\documentclass[vecphys]{svmult}

\usepackage{makeidx}         
\usepackage{graphicx}        
\usepackage{multicol}        
\usepackage[bottom]{footmisc}

\makeindex             

\def\lsim{\hbox{ \raise.35ex\rlap{$<$}\lower.6ex\hbox{$\sim$}\ }}
\def\gsim{\hbox{ \raise.35ex\rlap{$>$}\lower.6ex\hbox{$\sim$}\ }}

\def\xrightarrow#1#2#3#4{\,\lower#1pt\hbox{$\stackrel{\stackrel{\displaystyle #2}%
{\hbox to #3cm{\rightarrowfill}}}{#4}$}\,}
    
\newcommand{\Mpl}{M_{\rm Pl}}

\begin{document}

\title*{Production of Topological Defects at the End of Inflation}
\author{Mairi Sakellariadou}
\institute{Department of Physics, King's College, University of
London, Strand, London WC2R 2LS, United Kingdom.
\ \ \ 
\texttt{[email: Mairi.Sakellariadou@kcl.ac.uk\ ]}}

\maketitle

Cosmological inflation and topological defects have been considered
for a long time, either in disagreement or in competition.  On the one
hand an inflationary era is required to solve the shortcomings of the
hot big bang model, while on the other hand cosmic strings and
string-like objects are predicted to be formed in the early
universe. Thus, one has to find ways so that both can coexist. I
discuss how to reconcile cosmological inflation with cosmic strings.

\section{Introduction}
\label{sec:1}
For a number of years, inflation and cosmological defects have been
considered either as two incompatible or as two competing aspects of
modern cosmology. Let me explain why. Historically, one of the reasons
for which inflation was proposed is to rescue the standard hot big
bang model from the monopole problem. More precisely, setting an
inflationary era after the formation of monopoles, these unwanted
defects would have been diluted away. However, such a mechanism could
also dilute cosmic strings unless they were produced at the end or
after inflation. Later on, inflation and topological defects competed
as the two alternative mechanisms to provide the generation of density
perturbations leading to the observed large-scale structure and the
anisotropies in the Cosmic Microwave Background (CMB). However, the
inconsistency between predictions from topological defect models and
CMB data on the one hand, and the good agreement between adiabatic
fluctuations generated by the amplification of the quantum
fluctuations of the inflaton field on the other hand, indicated a
clear preference for inflation. Finally, the genericity of cosmic
string formation in the framework of Grand Unified Theories (GUTs) and
the formation of defect-like objects in brane cosmologies, convinced
us that cosmic strings have to play a r\^ole, which may be
sub-dominant but it is definitely there. This conclusion led to the
consideration of mixed models, where inflation and cosmic strings
coexist. The study of such models, the comparison of their
predictions against current data, and the consequences for the
theories within which we based our study is the aim of this study.

In Section \ref{sec:Inf}, I briefly describe cosmological inflation,
its success and its open questions. I then discuss hybrid inflation in
general and then I focus on F-/D-term inflation in the framework of
supersymmetry and supergravity theories. In Section \ref{sec:TD}, I
discuss topological defects in general, and cosmic strings in
particular. I then argue the genericity of string formation in the
framework of GUTs. In Section \ref{sec:bc}, I briefly discuss
braneworld cosmology, focusing on inflation within braneworld
cosmologies and the generation of cosmic superstrings. In Section
\ref{sec:oc}, I discuss observational consequences, and in particular
the spectrum of CMB anisotropies and that of gravity waves. I compare
the predictions of the models against current data, which allow me to
constrain the parameters space of the models. I round up with the
conclusions in Section \ref{sec:concl}.

\section{Cosmological Inflation}
\label{sec:Inf}
Despite its success, the standard hot big bang cosmological model has
a fairly severe drawback, namely the requirement, up to a high degree
of accuracy, of an initially homogeneous and flat universe. An
appealing solution to this problem is to introduce, during the very
early stages of the evolution of the universe, a period of accelerated
expansion, known as cosmological inflation \cite{infl}. The
inflationary era took place when the universe was in an unstable
vacuum-like state at a high energy density, leading to a
quasi-exponential expansion. The combination of the
hot big bang model and the inflationary scenario provides at present
the most comprehensive picture of the universe at our
disposal. Inflation ends when the Hubble parameter
$H=\sqrt{8\pi\rho/(3 M_{\rm Pl}^2)}$ (where $\rho$ denotes the energy
density and $M_{\rm Pl}$ stands for the Planck mass) starts decreasing
rapidly. The energy stored in the vacuum-like state gets transformed
into thermal energy, heating up the universe and leading to the
beginning of the standard hot big bang radiation-dominated era.

Inflation is based on the basic principles of general relativity and
field theory, while when the principles of quantum mechanics are also
considered, it provides a successful explanation for the origin of the
large scale structure, associated with the measured temperature
anisotropies in the CMB spectrum. Inflation is overall a very
successful scenario and many different models have been proposed and
studied over the last 25 years.  Nevertheless, inflation
still remains a paradigm in search of model. In principle, one should
search for an inflationary model inspired from some fundamental theory
and subsequently test its predictions against current data.  Moreover,
releasing the present universe form its acute dependence on the
initial data, inflation is faced with the challenging task of proving
itself generic, in the sense that inflation would take place without
fine-tuning of the initial conditions.  This issue, already addressed
in the past \cite{gp-cs}, has been recently re-investigated
\cite{GT-us}.

\subsection{Hybrid Inflation in SUSY GUTs}
\label{subsec:hisg}
Chaotic inflation \cite{chaotic} is, to my opinion, the most elegant
inflationary model. Nevertheless, in order for density inhomogeneities
generated at the end of inflation to have the required amplitude
$(\delta\rho/\rho)\sim 10^{-4}-10^{-5}$, the model requires fine-tuning.
In the simplest theory of a single scalar field minimally coupled to
gravity, the coupling must be of the order of
$\lambda\sim10^{-13}-10^{-14}$; the same fine-tuning was required in the
new inflationary model. This is a reason for which hybrid inflation
\cite{hybrid} has been proposed.

\begin{figure}
\centering
\includegraphics[height=4cm]{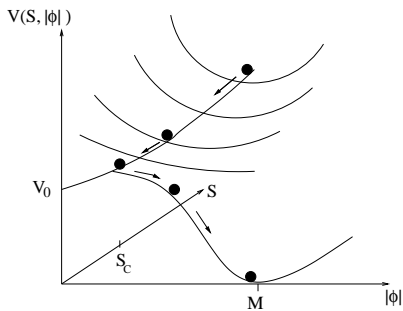}
\caption{A simplistic drawing of hybrid inflation.}
\label{fig:fig1}       
\end{figure}

Hybrid inflation is based on Einstein's gravity but is driven by false
vacuum. The inflaton field rolls down its potential while another
scalar field is trapped in an unstable false vacuum. Once the inflaton
field becomes much smaller than some critical value, a phase
transition to the true vacuum takes place and inflation ends (for an
illustration see Fig.~(\ref{fig:fig1})). Such a phase transition
may leave behind topological defects as false vacuum remnants. In
particular, the formation of topological defects may provide the
mechanism to gracefully exit the inflationary era in a number of
particle physics motivated inflationary models \cite{infldef}.

Theoretically motivated inflationary models can be built in the
context of supersymmetry or supergravity.  $N=1$ supersymmetry models
contain complex scalar fields which often have flat directions in
their potential, thus offering natural candidates for inflationary
models. In this framework, hybrid inflation driven by F-terms or
D-terms is the standard inflationary model, leading \cite{jrs}
generically to cosmic string formation at the end of inflation.
F-term inflation is potentially plagued with the $\eta$-problem, while
D-term inflation avoids it. Let me briefly explain what this problem
is. It is difficult to achieve slow-roll inflation within
supergravity, however inflation should last long enough to solve the
shortcomings of the standard big bang model. The positive false vacuum
of the inflaton field breaks spontaneously global supersymmetry, which
gets restored once inflation has been completed. However, since in
supergravity theories, supersymmetry breaking is transmitted by
gravity, all scalar fields acquire an effective mass of the order of
the expansion rate during inflation. Such a heavy mass for the scalar
field playing the r\^ole of the inflaton spoils the slow-roll
condition. It has been shown \cite{bd} that the {\sl Hubble-induced}
mass problem has its origin on the F-term interactions, while it
disappears if the vacuum energy is instead dominated by the D-terms of
the superfields.

\subsubsection{F-term Inflation}
F-term inflation can be naturally accommodated in the framework of
GUTs when a GUT gauge group, G$_{\rm GUT}$,
is broken down to the Standard Model (SM) gauge group, G$_{\rm SM}$,
at an energy scale $M_{\rm GUT}$ according to the scheme
\begin{equation}
\label{ssbF}
{\rm G}_{\rm GUT} \stackrel{M_{\rm GUT}}{\hbox to 0.8cm
{\rightarrowfill}} {\rm H}_1 \xrightarrow{9}{M_{\rm
infl}}{1}{\Phi_+\Phi_-} {\rm H}_2 {\longrightarrow} {\rm G}_{\rm SM}~;
\end{equation}
$\Phi_+, \Phi_-$ is a pair of GUT Higgs superfields in non-trivial
complex conjugate representations, which lower the rank of the group
by one unit when acquiring non-zero vacuum expectation value. The
inflationary phase takes place at the beginning of the symmetry
breaking ${\rm H}_1\stackrel{M_{\rm infl}}{\longrightarrow} {\rm
H}_2$.
The gauge symmetry is spontaneously broken by adding F-terms to the
superpotential. The Higgs mechanism leads generically \cite{jrs} to
Abrikosov-Nielsen-Olesen strings, called F-term strings.

F-term inflation is based on the globally supersymmetric
renormalisable superpotential
\begin{equation}\label{superpot}
W_{\rm infl}^{\rm F}=\kappa  S(\Phi_+\Phi_- - M^2)~,
\end{equation}
where $S$ is a GUT gauge singlet left handed superfield and $\kappa$, $M$
 are two constants ($M$ has dimensions of mass) which can be taken
 positive with field redefinition.  The scalar potential, as a function
 of the scalar complex component of the respective chiral superfields
 $\Phi_\pm, S$, reads
\begin{equation}
\label{scalpot1}
V(\phi_+,\phi_-, S)= |F_{\Phi_+}|^2+|F_{\Phi_-}|^2+|F_ S|^2
+\frac{1}{2}\sum_a g_a^2 D_a^2~.
\end{equation}
The F-term is such that $F_{\Phi_i} \equiv |\partial W/\partial
\Phi_i|_{\theta=0}$, where we take the scalar component of the
superfields once we differentiate with respect to $\Phi_i=\Phi_\pm,
 S$. The D-terms are
$D_a=\bar{\phi}_i\,{(T_a)^i}_j\,\phi^j +\xi_a$,
with $a$ the label of the gauge group generators $T_a$, $g_a$ the
gauge coupling, and $\xi_a$ the Fayet-Iliopoulos term. By
definition, in the F-term inflation the real constant $\xi_a$ is zero;
it can only be nonzero if $T_a$ generates an extra U(1) group.  In the
context of F-term hybrid inflation the F-terms give rise to the
inflationary potential energy density while the D-terms are flat
along the inflationary trajectory, thus one may neglect them during
inflation.
\begin{figure}
\centering
\includegraphics[height=4cm]{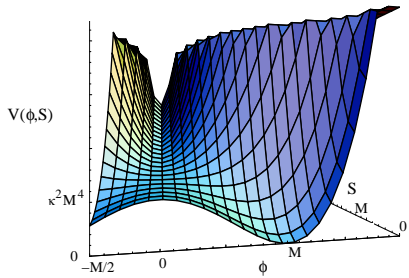}
\caption{A representation of the potential for F-term inflation in the
context of supersymmetry.}
\label{fig:fig2}       
\end{figure}

The potential, plotted in Fig.~(\ref{fig:fig2}), has one valley of
local minima, $V=\kappa^2 M^4$, for $S> M $ with $\phi_+ = \phi_-=0$,
and one global supersymmetric minimum, $V=0$, at $S=0$ and $\phi_+ =
\phi_- = M$. Imposing initially $ S \gg M$, the fields quickly settle
down the valley of local minima.  Since in the slow-roll inflationary
valley the ground state of the scalar potential is non-zero,
supersymmetry is broken.  In the tree level, along the inflationary
valley the potential is constant, therefore perfectly flat. A slope
along the potential can be generated by including one-loop radiative
corrections, which can be calculated using the Coleman-Weinberg
expression \cite{cw}
\begin{equation}\label{cw}
\Delta V_{1-{\rm loop}}=\frac{1}{64\pi^2}\sum_i (-1)^{F_i}
m_i^4\ln\frac{m_i^2}{\Lambda^2}~,
\end{equation}
where the sum extends over all helicity states i, with fermion number
$F_i$ and mass squared $m_i^2$; $\Lambda$ stands for a renormalisation
scale.  In this way, the scalar potential gets a little tilt which
helps the inflaton field $S$ to slowly roll down the valley of
minima. The one-loop radiative corrections to the scalar potential
along the inflationary valley lead to the effective potential
\cite{rs1}
\begin{eqnarray}
\label{VexactF}
V_{\rm eff}^{\rm F}(|S|)&=&\kappa^2M^4\biggl\{1+\frac{\kappa^2
\cal{N}}{32\pi^2}\biggl[2\ln\frac{|S|^2\kappa^2}{\Lambda^2}
+(z+1)^2
\ln(1+z^{-1})\nonumber\\
&&~~~~~~~~~+(z-1)^2\ln(1-z^{-1})
\biggr]\biggr\} ~~\mbox{with} ~~z=\frac{|S|^2}{M^2}~;
\end{eqnarray}
$\cal{N}$ stands for the dimensionality of the representation to which
the complex scalar components $\phi_+, \phi_-$ of the chiral
superfields $\Phi_+, \Phi_-$ belong. This implies that the effective
potential, Eq.~(\ref{VexactF}), depends on the particular symmetry
breaking scheme considered (see, Eq.~(\ref{ssbF})).

\subsubsection{D-term Inflation}
D-term inflation is one of the most interesting models of
inflation. It is possible to implement it naturally within high energy
physics, as for example Supersymmetric GUTS (SUSY GUTs), Supergravity
(SUGRA), or string theories. Moreover, it avoids the {\sl
Hubble-induced mass} problem. In D-term inflation, the gauge symmetry
is spontaneously broken by introducing Fayet-Iliopoulos (FI)
D-terms. In standard D-term inflation, the constant FI term gets
compensated by a single complex scalar field at the end of the
inflationary era, which implies that standard D-term inflation ends
with the formation of cosmic strings, called D-strings.  More
precisely, in its simplest form, the model requires a symmetry
breaking scheme
\begin{equation}
{\rm G}_{\rm GUT}\times {\rm U}(1) \stackrel{M_{\rm GUT}}{\hbox to
  0.8cm{\rightarrowfill}} {\rm H} \times {\rm U}(1)
\xrightarrow{9}{M_{\rm infl}}{1}{\Phi_+\Phi_-} {\rm H} \rightarrow
	    {\rm G}_{\rm SM}~.
\end{equation}

A supersymmetric description of the standard D-term inflation is
insufficient; the inflaton field reaches values of the order of the
Planck mass, or above it, even if one concentrates only around the
last 60 e-folds of inflation; the correct analysis is therefore in the
context of supergravity.

D-term inflation is based on the superpotential
\begin{equation}\label{superpoteninflaD}
W=\lambda S\Phi_+\Phi_-~,
\end{equation}
where $S, \Phi_+, \Phi_-$ are three chiral superfields and $\lambda$
is the superpotential coupling. In its standard form, the model
assumes an invariance under an Abelian gauge group $U(1)_\xi$, under
which the superfields $S, \Phi_+, \Phi_-$ have charges $0$, $+1$ and
$-1$, respectively. It is also assumed the existence of a constant
Fayet-Iliopoulos term $\xi$.  

In the \emph{standard} supergravity formulation the Lagrangian depends
on the K\"ahler potential $K(\Phi_i,\bar{\Phi}_i)$ and the
superpotential $W(\Phi_i)$ only through the combination
\begin{equation}
\label{kwcomb}
G(\Phi_i,\bar{\Phi}_i)= \frac{K(\Phi_i,\bar{\Phi}_i)}{\Mpl^2} +\ln
\frac{|W(\Phi_i)|^2}{\Mpl^6}~.
\end{equation}
However, this \emph{standard} supergravity formulation is
inappropriate to describe D-term inflation \cite{toine1}. In D-term
inflation the superpotential vanishes at the unstable de Sitter vacuum
(anywhere else the superpotential is nonzero). Thus, \emph{standard}
supergravity is inappropriate, since it is ill-defined at $W=0$. In
conclusion, D-term inflation must be described with a non-singular
formulation of supergravity when the superpotential vanishes.

Various formulations of effective supergravity can be constructed from
the superconformal field theory. One must first build a Lagrangian
with full superconformal theory, and then the gauge symmetries that
are absent in Poincar\'e supergravity must be gauge fixed. In this
way, one can construct a non-singular theory at $W=0$, where the
action depends on all three functions: the K\"ahler potential
$K(\Phi_i,\bar{\Phi}_i)$, the superpotential $W(\Phi_i)$ and the
kinetic function $f_{ab}(\Phi_i)$ for the vector multiplets.  To
construct a formulation of supergravity with constant Fayet-Iliopoulos
terms from superconformal theory, one finds \cite{toine1} that under
U(1) gauge transformations in the directions in which there are
constant Fayet-Iliopoulos terms $\xi_\alpha$, the superpotential $W$
must transform as \cite{toine1}
\begin{equation}
\delta_\alpha W=\eta_{\alpha i}\partial^i W = -i
\frac{g\xi_\alpha}{\Mpl^2}W~;
\end{equation}
it is incorrect to keep the same charge assignments as in
standard supergravity.

D-term inflationary models can be built with different choices of
K\"ahler geometry.  Let us first consider D-term inflation within minimal
supergravity. It is based on
\begin{equation}\label{Kmin}
K_{\rm min}=\sum_i |\Phi_i|^2=|\Phi_-|^2+|\Phi_+|^2+|S|^2~,
\end{equation}
with $f_{ab}(\Phi_i)=\delta_{ab}$. 
The tree level scalar potential is \cite{toine1}
\begin{eqnarray}\label{DpotenSUGRAtotbis}
V_{\rm min}=&&
\lambda^2\exp\left({\frac{|\phi_-|^2+|\phi_+|^2+|S|^2}{M^2_{\rm
Pl}}}\right)\nonumber 
\Biggl[|\phi_+\phi_-|^2\left(1+\frac{|S|^4}{M^4_{\rm
Pl}}\right)\nonumber\\
&&~~~~~~~~~~+|\phi_+S|^2 \left(1+\frac{|\phi_-|^4}{M^4_{\rm
Pl}}\right)
+|\phi_-S|^2 \left(1+\frac{|\phi_+|^4}{M^4_{\rm
Pl}}\right) +3\frac{|\phi_-\phi_+S|^2}{M^2_{\rm Pl}}\Biggr]\nonumber\\
&&+\ \frac{g^2}{2}\left(q_+|\phi_+|^2+q_-|\phi_-|^2+\xi\right)^2~,
\end{eqnarray}
with 
\begin{equation}
q_\pm = \pm 1-\xi/(2\Mpl^2)~.  
\end{equation} 
The potential has two minima: One global minimum at zero and one local
minimum equal to $V_0=(g^2/2)\xi^2$.  For arbitrary large $S$ the tree
level value of the potential remains constant and equal to $V_0$; the
$S$ plays the r\^ole of the inflaton field. Assuming chaotic initial
conditions $|S|\gg S_{\rm s}$, inflation begins. Along the
inflationary trajectory the D-term, which is the dominant one, splits
the masses in the $\Phi_\pm$ superfields, leading to the one-loop
effective potential for the inflaton field.  Considering the one-loop
radiative corrections \cite{rs1,prl2005}
\begin{equation}\label{scalarpeff}
V^{\rm eff}_{\rm min}(|S|)=\frac{g^2\xi^2}{2}\left\{
1+\frac{g^2}{16\pi^2}\left[2\ln\left(z\frac{g^2\xi}{\Lambda^2}\right)+f_V(z)
\right] \right\} ~,
\end{equation}
where
\begin{equation}
f_V(z) = (z+1)^2\ln\left( 1+\frac{1}{z}\right) + (z-1)^2\ln\left(
1-\frac{1}{z}\right)~,
\end{equation}
with
\begin{equation}
z\equiv \frac{\lambda^2}{g^2\xi} |S|^2
\exp\left(\frac{|S|^2}{M_{\rm Pl}^2}\right)~.
\end{equation}

As a second example, consider D-term inflation based on K\"ahler
geometry with a shift symmetry, $\phi\rightarrow\phi +c$ (where $c$ is
a real constant). Such models can lead \cite{shift} to flat enough
potentials with stabilisation of the volume of the compactified space.
They can therefore be used to built successful inflationary models
in the framework of string theories. The K\"ahler
potential is
\begin{equation}\label{K3}
K_{\rm shift}=\frac{1}{2} (S+\bar{S})^2+|\phi_+|^2+|\phi_-|^2~;
\end{equation}
the kinetic function has the minimal structure. The scalar
potential reads \cite{rs3}
\begin{eqnarray}
V_{\rm shift}&\simeq&
~\frac{g^2}{2}\left(|\phi_+|^2-|\phi_-|^2+\xi\right)^2 \nonumber\\
&&+\lambda^2\exp\left({\frac{|\phi_-|^2+|\phi_+|^2}{M^2_{\rm
Pl}}}\right)\exp\left[{\frac{(S+\bar{S})^2}{2M^2_{\rm Pl}}}\right]
\nonumber\\ & & ~~~\times
\Biggl[|\phi_+\phi_-|^2\left(1+\frac{S^2+\bar{S}^2}{M^2_{\rm
Pl}}+\frac{|S|^2|S+\bar{S}|^2}{M^4_{\rm Pl}}\right)+|\phi_+S|^2
\left(1+\frac{|\phi_-|^4}{M^4_{\rm Pl}}\right) \nonumber\\ &&
+|\phi_-S|^2 \left(1+\frac{|\phi_+|^4}{M^4_{\rm Pl}}\right)
+3\frac{|\phi_-\phi_+S|^2}{M^2_{\rm Pl}}\Biggr] ~.
\end{eqnarray}
As in D-term inflation within minimal supergravity, the potential has
a global minimum at zero for $\langle\Phi_+\rangle=0$ and
$\langle\Phi_-\rangle=\sqrt{\xi}$ and a local minimum equal to
$V_0=(g^2/2)\xi^2$ for $\langle S\rangle\gg S_{\rm c}$ and
$\langle\Phi_\pm\rangle=0$.  

The exponential factor $e^{|S|^2}$, which we got in the case of
minimal supergravity, has been replaced by $e^{(S+\bar{S})^2/2}$.
Writing $S=\eta+i\phi_0$ one gets $e^{(S+\bar{S})^2/2}=e^{\eta^2}$.
If $\eta$ plays the r\^ole of the inflaton field, we obtain the same
potential as for minimal D-term inflation. If instead $\phi_0$ is the
inflaton field, the inflationary potential is identical to that of the
usual D-term inflation within global supersymmetry \cite{rs1}.  The
latter case is better adapted with the choice $K_{\rm shift}$, since
then the exponential term is constant during inflation and thus it
cannot spoil the slow-roll conditions.

As a last example, consider a K\"ahler potential with
non-renormalisable terms:
\begin{eqnarray}
K_{\rm
non-renorm}&=&|S|^2+|\Phi_+|^2+|\Phi_-|^2\nonumber\\
&&+f_+\bigg(\frac{|S|^2}{M_{\rm
Pl}^2}\bigg)|\Phi_+|^2+f_-\bigg(\frac{|S|^2}{M_{\rm Pl}^2}\bigg)
|\Phi_-|^2+b\frac{|S|^4}{\Mpl^2}~,
\label{gen}
\end{eqnarray}
where $f_\pm$ are arbitrary functions of $(|S|^2/M_{\rm Pl}^2)$ and
the superpotential is given in Eq.~(\ref{superpoteninflaD}).  The
effective potential reads \cite{rs3}
\begin{equation}
V^{\rm eff}_{\rm non-renorm}(|S|)=\frac{g^2\xi^2}{2}\left\{
1+\frac{g^2}{16\pi^2}\left[ 2\ln \left(
z\frac{g^2\xi}{\Lambda^2}\right)+f_V(z) \right] \right\} ~,
\end{equation}
where 
\begin{equation}
\label{deffV}
f_V(z) = (z+1)^2\ln\left(1+\frac{1}{z}\right) + (z-1)^2\ln\left(
1-\frac{1}{z}\right)~
\end{equation}
\begin{equation}
\mbox{with} ~~~~z\equiv
\frac{\lambda^2|S|^2}{g^2\xi}\exp\bigg(\frac{|S|^2}{\Mpl^2}
+b\frac{|S|^4}{\Mpl^4}\bigg) \frac{1}{(1+f_+)(1+f_-)}~.
\end{equation}
The cosmological consequences of these inflationary models will be
presented in Section \ref{sec:oc}.

\section{Topological Defects in GUTs}
\label{sec:TD}
Following the standard version of the hot big bang model, the universe
could have expanded from a very hot (with a temperature $T\gsim
10^{19} {\rm GeV}$) and dense state, cooling towards its present
state.  As the universe expands and cools down, it undergoes a number
of phase transitions, breaking the symmetry between the different
interactions. Such phase transitions may leave behind topological
defects ~\cite{td} as false vacuum remnants, via the Kibble mechanism
\cite{kibble}.  Whether or not topological defects are formed during
phase transitions followed by Spontaneously Broken Symmetries (SSB)
depend on the topology of the vacuum manifold ${\cal M}_n$, which also
determines the type of the produced defects.  The properties of ${\cal
M}_n$ are usually described by the $k^{\rm th}$ homotopy group
$\pi_k({\cal M}_n)$, which classifies distinct mappings from the
$k$-dimensional sphere $S^k$ into the manifold ${\cal M}_n$.

Let me consider the symmetry breaking of a group G down to a subgroup
H of G . If ${\cal M}_n={\rm G}/{\rm H}$ has disconnected components,
or equivalently if the order $k$ of the nontrivial homotopy group is
$k=0$, two-dimensional defects, {\sl domain walls}, get formed.  The
spacetime dimension, $d$, of the defects is determined by the order of
the nontrivial homotopy group by $d=4-1-k$. If ${\cal M}_n$ is not
simply connected, meaning that ${\cal M}_n$ contains loops which
cannot be continuously shrunk into a point, {\sl cosmic strings} get
produced. A necessary but not sufficient condition for the formation
of stable strings is that the first (fundamental) homotopy group
$\pi_1({\cal M}_n)$ of ${\cal M}_n$, is nontrivial, or multiply
connected. Cosmic strings are line-like $(d=2)$ defects. If ${\cal
M}_n$ contains unshrinkable surfaces, then {\sl monopoles} $(k=1,
~d=1)$ get formed.  Finally, if ${\cal M}_n$ contains non-contractible
three-spheres, then event-like defects, called {\sl textures}, $(k=3,
~d=0)$ arise.

Depending on whether the original symmetry is local (gauged) or global
(rigid), topological defects are called local or global. The energy of
local defects is strongly confined, while the gradient energy of
global defects is spread out over the causal horizon at defect
formation.  Patterns of symmetry breaking which lead to the formation
of local monopoles or local domain walls are ruled out, since they
should soon dominate the energy density of the universe and close it,
unless an inflationary era took place after their formation.  Local
textures are insignificant in cosmology since their relative
contribution to the energy density of the universe decreases rapidly
with time \cite{textures}.

Even if the non-trivial topology required for the existence
of a defect is absent in a field theory, it may still be possible to have
defect-like solutions. Defects may be {\sl embedded}
in such topologically trivial field theories \cite{embedded}. While
stability of topological defects is guaranteed by topology, embedded
defects are in general unstable under small perturbations.

\subsection{Cosmic Strings}
\label{subsec:CS}
Cosmic strings \cite{mslnp} are analogous to flux tubes in type-II
superconductors, or to vortex filaments in superfluid helium.
Topologically stable strings do not have ends; they either form closed
loops or they extend to infinity. The linear mass density of strings,
$\mu$, which in the simplest models it also determines the string tension,
specifies the energy scale, $\eta$, of the symmetry breaking, $\mu\sim
\eta^2$. The strength of gravitational interactions of strings is
expressed in terms of the dimensionless parameter $G\mu\sim
\eta^2/M_{\rm Pl}^2$ (with $G$ the gravitational Newton's constant and
$M_{\rm Pl}$ the Planck mass).  For grand unification strings, the
energy per unit length is $\mu\sim 10^{22} {\rm kg}/{\rm m}$, or
equivalently, $G\mu\sim {\cal O}(10^{-6})$.

At formation, cosmic strings form a tangled network, made of Brownian
infinitely long strings and a distribution of closed loops. Curved
segments of strings moving under their tension reach almost
relativistic speeds. When two string segments intersect, they exchange
partners ({\sl intercommute}) with probability equal to 1.
String-string and self-string intersections lead to daughter
infinitely long strings and closed loops, as they can be seen in
Fig.~({\ref{fig:fig3}). 
\begin{figure}
\centering
\includegraphics[height=3cm]{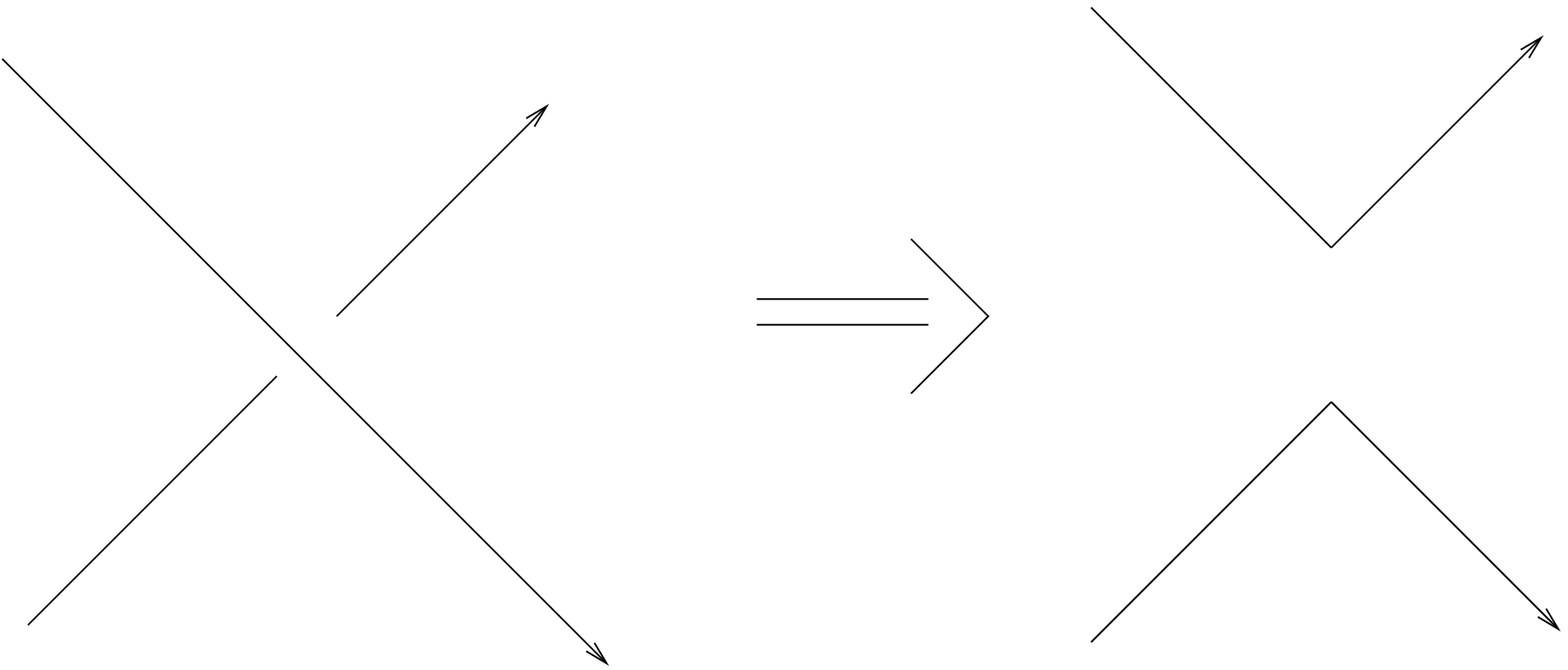}
\vskip.8cm
\includegraphics[height=3cm]{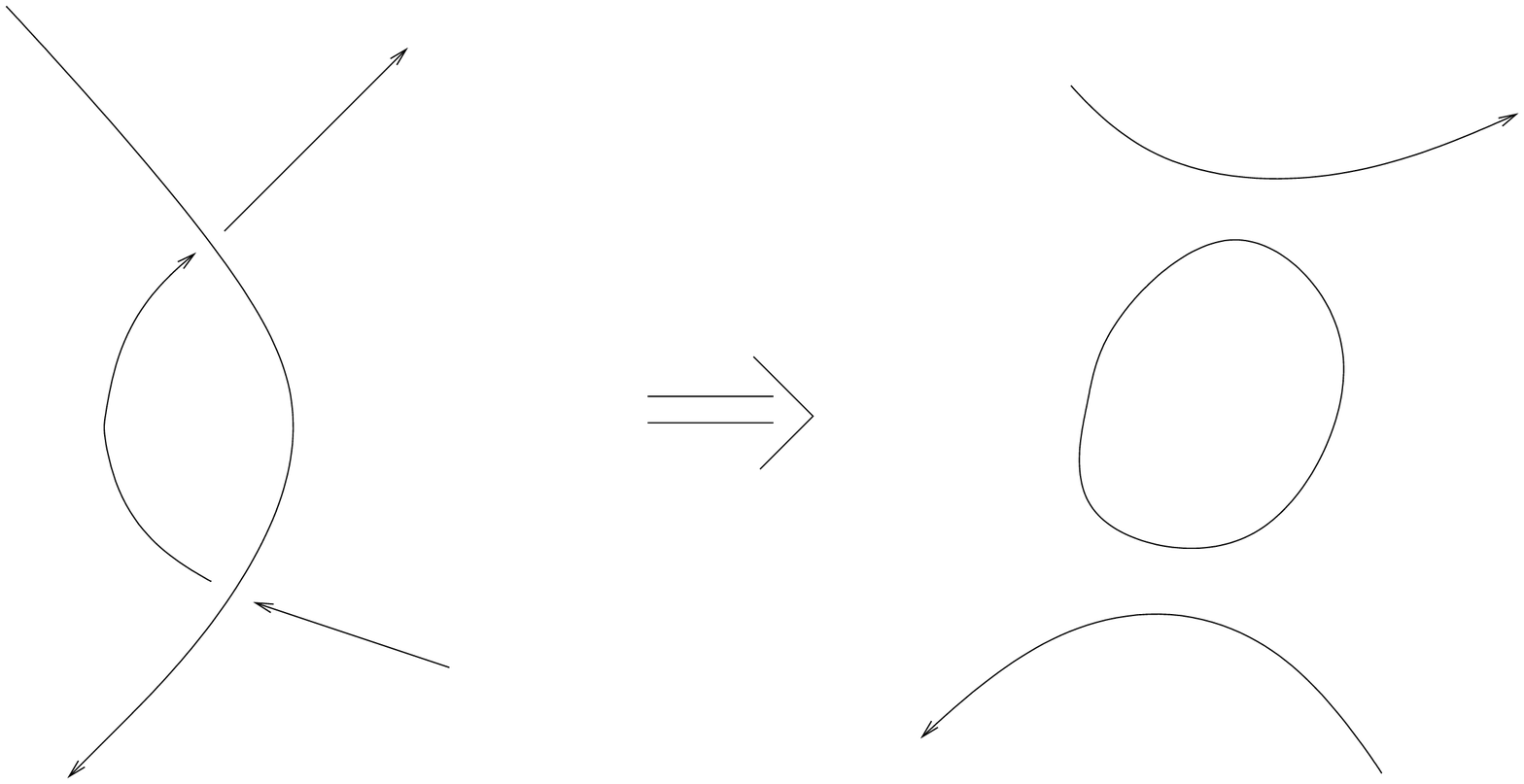}
\vskip.8cm
\includegraphics[height=3cm]{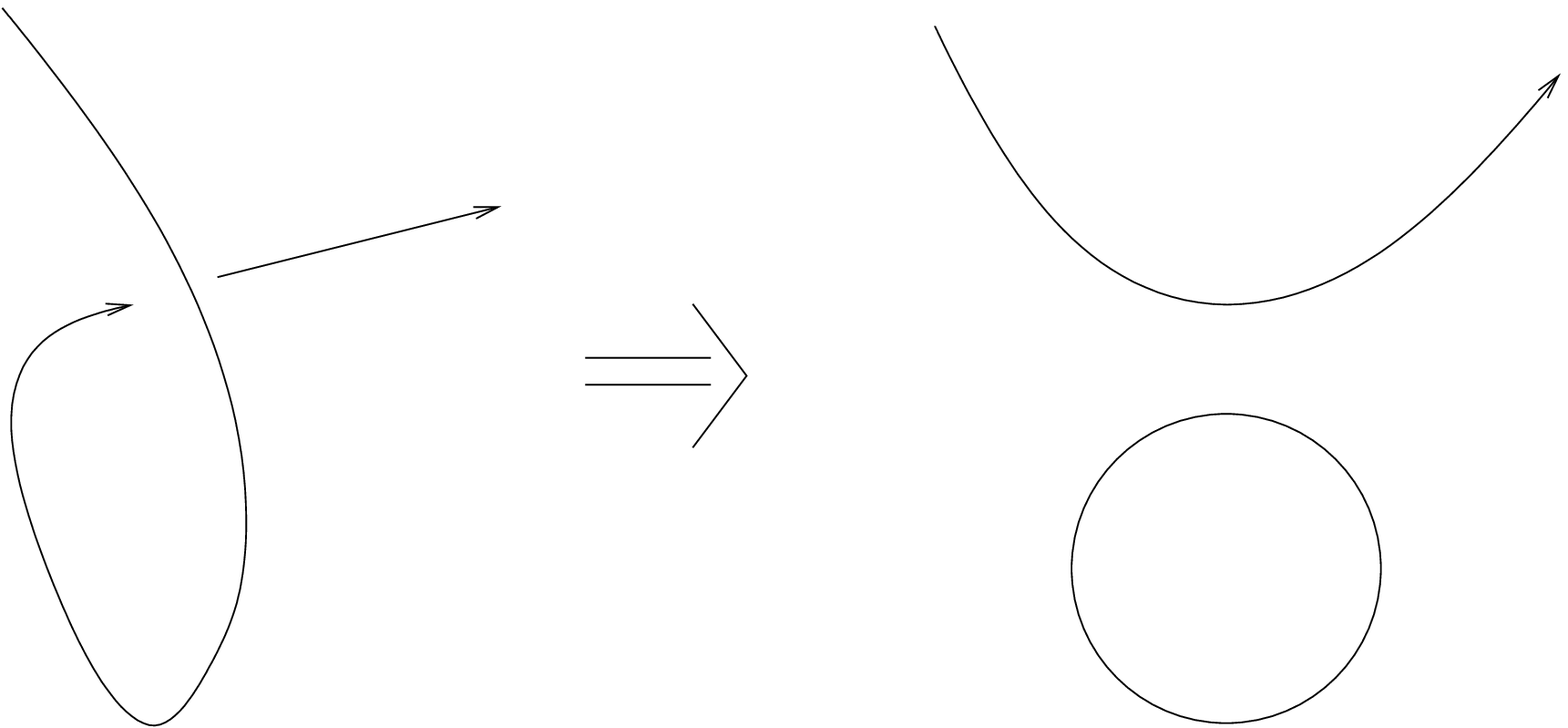}
\caption{At the top, string-string interactions at one point leading
to the formation of two new long strings via exchange of partners. In
the middle, string-string interactions at two points, leading to two
new long strings and a loop. At the bottom, self-self interactions
leading to the formation of a new long string and a loop
\cite{mslnp}.}
\label{fig:fig3}       
\end{figure}
Clearly, string intercommutations produce discontinuities on the new
string segments at the intersection point. These discontinuities ({\sl
kinks}) are composed of right- and left-moving pieces travelling along
the string at the speed of light.

Early analytic work \cite{one-scale} identified the key property of
{\sl scaling}, where at least the basic properties of the string
network can be characterised by a single length scale, roughly the
persistence length (defined as the distance beyond which the
directions along the string are uncorrelated), $\xi(t)$, and the
typical separation between string segments, $d(t)$, both grow with the
cosmic horizon.  This result was supported by subsequent numerical
work \cite{numcs}.  However, further investigation revealed dynamical
processes, including loop production, at scales much smaller than
$\xi$ \cite{proc}.

Recent numerical simulations of cosmic string evolution in a expanding
universe found evidence \cite{cmf} of a scaling regime for the cosmic
string loops in the radiation and matter dominated eras down to the
hundredth of the horizon time. It is important to note that the
scaling was found without considering any gravitational back reaction
effect; it was just the result of string intercommuting mechanism. As
it was reported in Ref.~\cite{cmf}, the scaling regime of string loops
appears after a transient relaxation era, driven by a transient
overproduction of string loops with lengths close to the initial
correlation length of the string network. Calculating the amount of
energy momentum tensor lost from the string network, it was found
\cite{cmf} that a few percents of the total string energy density
disappear in the very brief process of formation of numerically
unresolved string loops during the very first timesteps of the string
evolution. Subsequently, other studies supported these findings
\cite{vov}.  A snapshot of the evolution of a cosmic string network
during the matter dominated era is shown in Fig.~\ref{fig:fig4}.
\begin{figure}
\centering
\includegraphics[height=4cm]{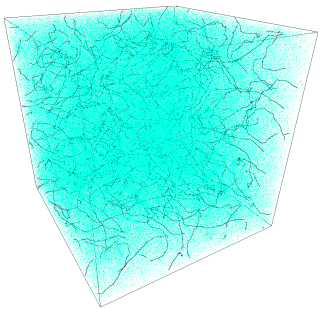}
\caption{Snapshot of a string network in the matter-dominated era
\cite{cmf}.}
\label{fig:fig4}       
\end{figure}

\subsection{Genericity of Cosmic String Formation within SUSY GUTs}
\label{subsec:gcs}
To investigate the cosmological consequences of cosmic strings formed
at the end of hybrid inflation, one should first address the question
of whether such objects are generically formed. I will briefly discuss
the genericity of cosmic string formation in the framework of SUSY
GUTS.

Even though the Standard Model has been tested to a very high
precision, it is incapable of explaining neutrino masses \cite{SK}.
An extension of the Standard Model gauge group can be realised within
Supersymmetry (SUSY). SUSY offers a solution to the gauge hierarchy
problem, while in the supersymmetric standard model the gauge coupling
constants of the strong, weak and electromagnetic interactions meet at
a single point $M_{\rm GUT} \simeq (2-3) \times 10^{16}$ GeV.  In
addition, SUSY GUTs can provide the scalar field which could drive
inflation, explain the matter-antimatter asymmetry of the universe,
and propose a candidate, the lightest superparticle, for cold dark
matter.

Within SUSY GUTs there is a large number of SSB patterns leading from
a large gauge group G to the SM gauge group G$_{\rm SM}\equiv$
SU(3)$_{\rm C}\times$ SU(2)$_{\rm L}\times$ U(1)$_{\rm Y}$. The study
of the homotopy group of the false vacuum for each SSB scheme will
determine whether there is defect formation and it will identify the
type of the formed defect. Clearly, if there is formation of domain
walls or monopoles, one will have to place an era of supersymmetric
hybrid inflation to dilute them. To consider a SSB scheme as a
successful one, it should be able to explain the matter/anti-matter
asymmetry of the universe and to account for the proton lifetime
measurements \cite{SK}.  In what follows, I consider a mechanism of
baryogenesis via leptogenesis, which can be thermal or non-thermal
one.  In the case of non-thermal leptogenesis, U(1)$_{\rm B-L}$ (B and
L, are the baryon and lepton numbers, respectively) is a sub-group of
the GUT gauge group, G$_{\rm GUT}$, and B-L is broken at the end or
after inflation. In the case of thermal leptogenesis, B-L is broken
independently of inflation. If leptogenesis is thermal and B-L is
broken before the inflationary era, then one should check whether the
temperature at which B-L is broken, which will define the mass of the
right-handed neutrinos, is smaller than the reheating temperature
which should be lower than the limit imposed by the gravitino. To
ensure the stability of proton, the discrete symmetry Z$_2$, which is
contained in U(1)$_{\rm B-L}$, must be kept unbroken down to low
energies. This implies that the successful SSB schemes should end at
G$_{\rm SM}\times$ Z$_2$.  I will then examine how often cosmic
strings have survived after the inflationary era, within all
acceptable SSB patterns.

To accomplish this task one has to choose the large gauge group
G$_{\rm GUT}$.  In Ref. \cite{jrs} this study has been done explicitly
for a large number of simple Lie groups. Since I consider GUTs based
on simple gauge groups, the type of supersymmetric hybrid inflation
will be of the F-type. The minimum rank of G$_{\rm GUT}$ has to be at
least equal to 4, to contain the G$_{\rm SM}$ as a subgroup.  Then one
has to study the possible embeddings of G$_{\rm SM}$ in G$_{\rm GUT}$
to be in agreement with the Standard Model phenomenology and
especially with the hypercharges of the known particles. Moreover, the
group must include a complex representation, needed to describe the
Standard Model fermions, and it must be anomaly free.  Since, in
principle, ${\rm SU}(n)$ may not be anomaly free, I assume that the
${\rm SU}(n)$ groups which I use, they have indeed a fermionic
representation that certifies that the model is anomaly free. I set as
the upper bound on the rank $r$ of the group, $r\leq 8$. Clearly, the
choice of the maximum rank is in principle arbitrary.  This choice
could, in a sense, be motivated by the Horava-Witten \cite{hw} model,
based on ${\rm E}_8\times {\rm E}_8$. Thus, the large gauge group
G$_{\rm GUT}$ could be one of the following: SO(10), E$_6$, SO(14),
SU(8), SU(9); flipped SU(5) and [SU(3)]$^3$ are included within this
list as subgroups of SO(10) and E$_6$, respectively.

A detailed study of all the SSB schemes which bring us from G$_{\rm
GUT}$ down to the Standard Model gauge group G$_{\rm SM}$, by one or more
intermediate steps, shows that cosmic strings are generically formed
at the end of hybrid inflation.  If the large gauge group G$_{\rm
GUT}$ is SO(10) then cosmic strings formation is unavoidable
\cite{jrs, rachel}.  For ${\rm
E}_6$ it depends whether one considers thermal or non-thermal
leptogenesis. More precisely, under the assumption of non-thermal
leptogenesis then cosmic strings formation is unavoidable. If I
consider thermal leptogenesis then cosmic strings formation at the end
of hybrid inflation arises in $98\%$ of the acceptable SSB schemes
\cite{jm}.  If the requirement of having Z$_2$ unbroken down to low
energies is relaxed and thermal leptogenesis is considered as being
the mechanism for baryogenesis, then cosmic strings formation
accompanies hybrid inflation in $80\%$ of the SSB schemes \cite{jm}.

For an illustration I give below the list of the SSB
schemes of ${\rm E}_6$ down to the G$_{\rm SM}\times Z_2$ via ${\rm
SO}(10)\times{\rm U}(1)$ (the reader is referred to Ref.~\cite{jrs}
for a full analysis).
Every $\stackrel{n}{\longrightarrow}$ represent an SSB
during which there is formation of topological defects, whose type
is denoted by $n$: $1$ for monopoles, $2$ for topological cosmic
strings, $2'$ for embedded strings, $3$ for domain walls. Note
that for e.g. $3_{\rm C}~2_{\rm L}~2_{\rm R}~1_{\rm B-L}$ stands
for SU(3)$_{\rm C}\times$ SU(2)$_{\rm L}\times$ SU(2) $_{\rm
R}\times$ U(1)$_{\rm B-L}$.  

\begin{equation}
    {\begin{array}{cllllll}
{\rm E}_6
\stackrel{1}{\rightarrow} {\rm SO(10)} ~1_{\rm V'}  &
  \left\{
    \begin{array}{cllllccccc}
 \stackrel{2}{\longrightarrow}  & {\rm SO(10)}  & \stackrel{}{\longrightarrow} ~~~\mbox{{\rm Eq.~(\ref{eq:51})}}\\
 \stackrel{1}{\longrightarrow} & 5  ~1_{\rm V}   ~1_{\rm V'}  & \stackrel{}{\longrightarrow} ~~~ \mbox{{\rm Eq.~(\ref{eq:5})}} \\
 \stackrel{1}{\longrightarrow}  &  5_{\rm F}   ~1_{\rm V}   ~1_{\rm V'}
& \stackrel{}{\longrightarrow} ~~~ \mbox{{\rm Eq.~(\ref{eq:5F})}}
\\
 \stackrel{1}{\longrightarrow}  &  5_{\rm E}   ~1_{\rm V}   ~1_{\rm V'}  & \stackrel{2',2}{\longrightarrow}   ~~{\rm G}_{\rm SM}  ~Z_2 \\
 \stackrel{2}{\longrightarrow} & 5  ~1_{\rm V'}  ~ Z_2   & \stackrel{}{\longrightarrow}   ~~~\mbox{{\rm Eq.~(\ref{eq:5}a)}} \\
 \stackrel{1,2}{\longrightarrow} & 5  ~1_{\rm V}  & \stackrel{}{\longrightarrow}  
 ~~~\mbox{{\rm Eq.~(\ref{eq:51}a)}}\\
  \stackrel{1}{\longrightarrow}  &  5_{\rm F}   ~1_{\rm V} & \stackrel{2',2}{\longrightarrow}  ~~{\rm G}_{\rm SM}  ~ Z_2 \\
\stackrel{1}{\longrightarrow} &  {\rm G}_{\rm SM}   ~1_{\rm V}  & \stackrel{2}{\longrightarrow}   ~~{\rm G}_{\rm SM}  ~Z_2 \\
 \stackrel{1,2}{\longrightarrow} &  {\rm G}_{\rm SM}   ~1_{\rm V'}  ~Z_2  & \stackrel{2}{\longrightarrow}   ~~{\rm G}_{\rm SM}~ Z_2 \\
\stackrel{1,2}{\longrightarrow}& 4_{\rm C} ~2_{\rm L} ~2_{\rm R} ~1_{\rm V'}
& \stackrel{}{\longrightarrow} ~~~ \mbox{{\rm Eq.~(\ref{eq:ps6})}}
\\
\stackrel{1}{\longrightarrow}& 4_{\rm C} ~2_{\rm L} ~2_{\rm R} & \stackrel{}{\longrightarrow} ~~~\mbox{{\rm Eq. (\ref{eq:ps})}}\\
\stackrel{1}{\longrightarrow}& 3_{\rm C} ~2_{\rm L} ~2_{\rm R} ~1_{\rm B-L} ~1_{\rm V'} & \stackrel{}{\longrightarrow} 
~~~ \mbox{{\rm Eq.~(\ref{eq:ps6}c)}}
\\
\stackrel{1}{\longrightarrow}& 3_{\rm C} ~2_{\rm L} ~1_{\rm R} ~1_{\rm B-L} ~1_{\rm V'} & \stackrel{}{\longrightarrow} 
~~~ \mbox{{\rm Eq.~(\ref{eq:ps6}b)}}
\\

  \end{array}
  \right.
    \end{array}  }
\end{equation}
where
\begin{equation}
\label{eq:51}
    {
  {\rm SO(10)} \left\{ \begin{array}{cccccc}
    \stackrel{1}{\longrightarrow} & 5   ~1_{\rm V} &
      \stackrel{1}{\longrightarrow}& 3_{\rm C} ~2_{\rm L} ~1_{\rm Z} ~1_{\rm V}
      &\stackrel{2}{\longrightarrow}& {\rm G}_{\rm SM}~Z_2\\
  \stackrel{1}{\longrightarrow}  &   4_{\rm C} ~2_{\rm L}  ~2_{\rm R}   &
\stackrel{}{\longrightarrow}    & \mbox{\rm Eq. (\ref{eq:ps})}
\\
      \stackrel{1,2}{\longrightarrow}  &   4_{\rm C} ~2_{\rm L}   ~2_{\rm R}     ~Z_2^{\rm C}   &  \stackrel{}{\longrightarrow}   
 & \mbox{{\rm Eq. (\ref{eq:psZ2})}}
\\
  \stackrel{1,2}{\longrightarrow}  &   4_{\rm C}     ~2_{\rm L}     ~1_{\rm R}     ~Z_2^{\rm C}   &  \stackrel{}{\longrightarrow}  
& \mbox{{\rm Eq. (\ref{eq:psZ2}b)}}\\
  \stackrel{1}{\longrightarrow}  &   4_{\rm C}     ~2_{\rm L}     ~1_{\rm R}   &  \stackrel{}{\longrightarrow}  &    
\mbox{\rm Eq. (\ref{eq:ps}b)}\\
  \stackrel{1,2}{\longrightarrow}  &   3_{\rm C}     ~2_{\rm L}     ~2_{\rm R}     ~1_{\rm B-L}     ~Z_2^{\rm C}   &  \stackrel{}{\longrightarrow} 
& \mbox{{\rm Eq. (\ref{eq:psZ2}a)}}\\
  \stackrel{1}{\longrightarrow}  &   3_{\rm C}     ~2_{\rm L}     ~2_{\rm R}     ~1_{\rm B-L}   &  \stackrel{}{\longrightarrow}   
& \mbox{{\rm Eq. (\ref{eq:ps}a)}}\\
  \stackrel{1}{\longrightarrow}  &   3_{\rm C}     ~2_{\rm L}     ~1_{\rm R}     ~1_{\rm B-L}   &  \stackrel{2}{\longrightarrow}   &   {\rm G}_{\rm SM}  ~Z_2  \\
  \end{array}
  \right.
    }
\end{equation}

\begin{equation}
\label{eq:5}
\begin{array}{clllllccc}
 5  ~1_{\rm V}   ~1_{\rm V'}
& \left\{
\begin{array}{clllllccc}
\stackrel{2}{\longrightarrow} & 5  ~1_{\rm V'}  ~Z_2
& \left\{
\begin{array}{clllllccc}
 \stackrel{1}{\longrightarrow} &  {\rm G}_{\rm SM}   ~1_{\rm V'}  ~ Z_2  & \stackrel{2}{\longrightarrow} &  {\rm G}_{\rm SM}  ~Z_2 \\
\end{array}
  \right.
\\
\stackrel{1}{\longrightarrow} &  {\rm G}_{\rm SM}   ~1_{\rm V}   ~1_{\rm V'}
& \left\{
\begin{array}{cllllllccc}
\stackrel{2}{\longrightarrow} &  {\rm G}_{\rm SM}   ~1_{\rm V}  & \stackrel{2}{\longrightarrow} &  {\rm G}_{\rm SM}  ~ Z_2  \\
 \stackrel{2}{\longrightarrow} &  {\rm G}_{\rm SM}   ~1_{\rm V'}  ~Z_2  & \stackrel{2}{\longrightarrow} &  {\rm G}_{\rm SM}  ~ Z_2 \\
\end{array}
  \right.
\\
 \stackrel{2}{\longrightarrow} & 5  ~1_{\rm V} & \stackrel{}{\longrightarrow}   
 ~~~\mbox{{\rm Eq.~(\ref{eq:51}a)}}\\
\end{array}
  \right.\end{array}
\end{equation}

\begin{equation}
\label{eq:5F}
\begin{array}{clllllccc}
 5_{\rm F}   ~1_{\rm V}   ~1_{\rm V'}
& \left\{
\begin{array}{cllllccc}
~\stackrel{2}{\longrightarrow} &  ~~5_{\rm F}   ~1_{\rm V}   & \stackrel{2',2}{\longrightarrow}   & {\rm G}_{\rm SM} ~Z_2 \\
 ~\stackrel{2',2}{\longrightarrow} &  ~{\rm G}_{\rm SM} ~Z_2 \\
\end{array}
  \right.
\end{array}
\end{equation}

\begin{equation}
\label{eq:ps6}
\begin{array}{clllllccc}
4_{\rm C} ~2_{\rm L} ~2_{\rm R} ~1_{\rm V'}
&\left\{
\begin{array}{cllllccc}
\stackrel{2}{\longrightarrow}& 4_{\rm C} ~2_{\rm L} ~2_{\rm R} & \stackrel{}{\longrightarrow} ~~~\mbox{{\rm Eq.~(\ref{eq:ps})}} \\
\stackrel{1}{\longrightarrow}& 3_{\rm C} ~2_{\rm L} ~1_{\rm R} ~1_{\rm B-L} ~1_{\rm V'}
& \left\{
\begin{array}{cllllllccc}
\stackrel{2}{\longrightarrow}& 3_{\rm C}  ~2_{\rm L} ~1_{\rm R} ~1_{\rm B-L} &\stackrel{2}{\longrightarrow}&  {\rm G}_{\rm SM}~Z_2\\
\stackrel{2',2}{\longrightarrow}& {\rm G}_{\rm SM} 1_{\rm V'}~Z_2 &\stackrel{2}{\longrightarrow}& {\rm G}_{\rm SM}~Z_2\\
\stackrel{2',2}{\longrightarrow}& {\rm G}_{\rm SM}~Z_2\\
  \end{array}
  \right.
\\
\stackrel{1}{\longrightarrow}& 3_{\rm C} ~2_{\rm L} ~2_{\rm R} ~1_{\rm B-L} ~1_{\rm V'}
& \left\{
\begin{array}{cllllccc}
\stackrel{1}{\longrightarrow}& 3_{\rm C} ~2_{\rm L} ~1_{\rm R} ~1_{\rm B-L} ~1_{\rm V'} &\stackrel{}{\longrightarrow}
~~~\mbox{{\rm Eq.~(\ref{eq:ps6}b)}} \\
\stackrel{2}{\longrightarrow}& 3_{\rm C} ~2_{\rm L} ~2_{\rm R} ~1_{\rm B-L}&\stackrel{}{\longrightarrow}
~~~\mbox{{\rm Eq.~(\ref{eq:ps}a)}} \\
\stackrel{2',2}{\longrightarrow}& {\rm G}_{\rm SM} 1_{\rm V'}~Z_2 &\stackrel{2}{\longrightarrow} ~~~{\rm G}_{\rm SM}~Z_2\\
\stackrel{1,2}{\longrightarrow}& 3_{\rm C} ~2_{\rm L} ~1_{\rm R} ~1_{\rm B-L} &\stackrel{2}{\longrightarrow} ~~~{\rm G}_{\rm SM}~Z_2\\
\stackrel{2',2}{\longrightarrow}& {\rm G}_{\rm SM}~Z_2\\
  \end{array}
  \right.
\\
\stackrel{1}{\longrightarrow}& 4_{\rm C} ~2_{\rm L} ~1_{\rm R} ~1_{\rm V'}
& \left\{
\begin{array}{cllllccc}
 \stackrel{2}{\longrightarrow}& 4_{\rm C} ~2_{\rm L} ~1_{\rm R}  &\stackrel{}{\longrightarrow}& 
~~~\mbox{{\rm Eq.~(\ref{eq:ps}b)}} \\
\stackrel{1}{\longrightarrow}& 3_{\rm C} ~2_{\rm L} ~1_{\rm R} ~1_{\rm B-L} ~1_{\rm V'}  &\stackrel{}{\longrightarrow}
&~~~\mbox{{\rm Eq.~(\ref{eq:ps6}b)}} \\
\stackrel{2',2}{\longrightarrow}& {\rm G}_{\rm SM} 1_{\rm V'}~Z_2 &\stackrel{2}{\longrightarrow}& {\rm G}_{\rm SM}~Z_2\\
\stackrel{1,2}{\longrightarrow}& 3_{\rm C} ~2_{\rm L} ~1_{\rm R} ~1_{\rm B-L} &\stackrel{2}{\longrightarrow}& {\rm G}_{\rm SM}~Z_2 \\
\stackrel{2}{\longrightarrow}& {\rm G}_{\rm SM}~Z_2\\
 \end{array}
  \right.
\\
\stackrel{1,2}{\longrightarrow}& {\rm G}_{\rm SM} 1_{\rm V'}~Z_2 & \stackrel{2}{\longrightarrow} ~~~{\rm G}_{\rm SM}~Z_2\\
\stackrel{1,2}{\longrightarrow}& 3_{\rm C} ~2_{\rm L} ~2_{\rm R} ~1_{\rm B-L} & \stackrel{}{\longrightarrow} 
~~~\mbox{{\rm Eq.~(\ref{eq:ps}a)}} \\
\stackrel{1}{\longrightarrow}& 3_{\rm C} ~2_{\rm L} ~1_{\rm R} ~1_{\rm B-L} & \stackrel{2}{\longrightarrow} ~~~{\rm G}_{\rm SM}~Z_2\\
  \end{array}
  \right.\end{array}
\end{equation}
with

 \begin{equation}
\label{eq:ps}
\begin{array}{clllcccc}
 4_{\rm C}     ~2_{\rm L}     ~2_{\rm R}   &
\left\{
\begin{array}{cllllccc}
\stackrel{1}{\longrightarrow} & 3_{\rm C} ~2_{\rm L} ~2_{\rm R} ~1_{\rm
B-L} & \left\{
\begin{array}{cllllccc}
 \stackrel{1}{\longrightarrow}  &   3_{\rm C}     ~2_{\rm L}     ~1_{\rm R}     ~1_{\rm B-L}   &  \stackrel{2}{\longrightarrow}  &   {\rm G}_{\rm SM}  ~Z_2   \\
  \stackrel{2',2}{\longrightarrow}  &   {\rm G}_{\rm SM} ~Z_2\\
 \end{array}
\right.
\\
  \stackrel{1}{\longrightarrow}  &   4_{\rm C}     ~2_{\rm L}     ~1_{\rm R}   &
\left\{
\begin{array}{cllllccc}
  \stackrel{1}{\longrightarrow}  &   3_{\rm C}     ~2_{\rm L}     ~1_{\rm R}     ~1_{\rm B-L}   &   \stackrel{2}{\longrightarrow}  &   {\rm G}_{\rm SM} ~Z_2\\
 \stackrel{2',2}{\longrightarrow}  &   {\rm G}_{\rm SM} ~Z_2\\
 \end{array}
\right.
\\
  \stackrel{1}{\longrightarrow}  &   3_{\rm C}     ~2_{\rm L}     ~1_{\rm R}     ~1_{\rm B-L}   &  ~~~~\stackrel{2}{\longrightarrow}      ~~{\rm G}_{\rm SM} ~Z_2\\
  \end{array}
\right.
\end{array}
\end{equation}

\begin{equation}
\label{eq:psZ2}
\begin{array}{clllcccc}
  4_{\rm C}     ~2_{\rm L}     ~2_{\rm R}     ~Z_2^{\rm C}   &
\left\{
\begin{array}{cllllccc}

 \stackrel{1 }{\longrightarrow}  &    3_{\rm C}     ~2_{\rm L}     ~2_{\rm R}     ~1_{\rm B-L}     ~Z_2^{\rm C}   &
\left\{
\begin{array}{cllllccc}
\stackrel{3}{\longrightarrow}  &    3_{\rm C}     ~2_{\rm L}     ~2_{\rm R}     ~1_{\rm B-L}   &   \stackrel{}{\longrightarrow}   &   
~~~\mbox{\rm Eq. (\ref{eq:ps}a)}   \\
 \stackrel{1,3}{\longrightarrow}  &   3_{\rm C}     ~2_{\rm L}     ~1_{\rm R}     ~1_{\rm B-L}   &  \stackrel{2}{\longrightarrow}  &   {\rm G}_{\rm SM}  ~ Z_2 \\
  \end{array}
\right.
\\
  \stackrel{1}{\longrightarrow}  &   4_{\rm C}     ~2_{\rm L}     ~1_{\rm R}     ~Z_2^{\rm C}   &
\left\{
\begin{array}{cllllccc}
  \stackrel{3}{\longrightarrow}  &   4_{\rm C}     ~2_{\rm L}     ~1_{\rm R}   &  {\longrightarrow}  &  
~~~\mbox{\rm Eq. (\ref{eq:ps}b)} \\
  \stackrel{1,3}{\longrightarrow}  &   3_{\rm C}     ~2_{\rm L}     ~1_{\rm R}     ~1_{\rm B-L}   &  \stackrel{2}{\longrightarrow}  &  {\rm G}_{\rm SM} ~Z_2\\
  \end{array}
\right.\\
\stackrel{3}{\longrightarrow}  &   4_{\rm C}     ~2_{\rm L}     ~2_{\rm R}
&{\longrightarrow}   ~~~\mbox{\rm Eq. (\ref{eq:ps})}   \\
  \stackrel{1}{\longrightarrow}  &   4_{\rm C}     ~2_{\rm L}     ~1_{\rm R}   &  {\longrightarrow}    
~~~\mbox{\rm Eq. (\ref{eq:ps}b)} \\
\stackrel{1,3}{\longrightarrow}  &   3_{\rm C}     ~2_{\rm L}     ~2_{\rm R}     ~1_{\rm B-L}   &  \stackrel{}{\longrightarrow}     ~
~~~\mbox{\rm Eq. (\ref{eq:ps}a)} \\
  \stackrel{1, 3}{\longrightarrow}  &   3_{\rm C}     ~2_{\rm L}     ~1_{\rm R}     ~1_{\rm B-L}    &  \stackrel{2}{\longrightarrow}     ~~~ {\rm G}_{\rm SM} ~Z_2
 \end{array}
\right.
\end{array}
\end{equation}
In addition, there are more direct schemes; they are listed below:

\begin{equation}
    {
{\rm E}_6\left\{ \begin{array}{clllllcccc}
&\stackrel{1}{\longrightarrow}& 5 ~1_{\rm V} ~1_{\rm V'} &\stackrel{}{\longrightarrow}& \mbox{{\rm Eq.(\ref{eq:5})}}\\
&\stackrel{1}{\longrightarrow}& 5_{\rm F} ~1_{\rm V} ~1_{\rm V'} &\stackrel{}{\longrightarrow}& \mbox{{\rm Eq.(\ref{eq:5F})}}\\
&\stackrel{1}{\longrightarrow}& 5_{\rm E} ~1_{\rm V} ~1_{\rm V'} &\stackrel{2',2}{\longrightarrow}& {\rm G}_{\rm SM}~Z_2\\
&\stackrel{1}{\longrightarrow}& 5 ~1_{\rm V}  &\stackrel{}{\longrightarrow}& 
\mbox{{\rm Eq.(\ref{eq:51}a)}}\\
&\stackrel{1}{\longrightarrow}& 5 ~1_{\rm V'}  &\stackrel{}{\longrightarrow}& 
\mbox{{\rm Eq.(\ref{eq:5}a)}}\\
&\stackrel{1}{\longrightarrow}& 5_{\rm F} ~1_{\rm V}  &\stackrel{2',2 }{\longrightarrow}& {\rm G}_{\rm SM} ~Z_2\\
&\stackrel{1}{\longrightarrow}& 4_{\rm C} ~2_{\rm L} ~2_{\rm R}
~1_{\rm V'} &\stackrel{}{\longrightarrow}& \mbox{\rm Eq. (\ref{eq:ps6})}\\
&\stackrel{1}{\longrightarrow}& 4_{\rm C} ~2_{\rm L} ~2_{\rm R}
&\stackrel{}{\longrightarrow}& \mbox{{\rm Eq.~(\ref{eq:ps})}}\\
&\stackrel{1}{\longrightarrow}& 4_{\rm C} ~2_{\rm L} ~1_{\rm R}&\stackrel{}{\longrightarrow}& 
\mbox{{\rm Eq.~(\ref{eq:ps}b)}}\\
&\stackrel{1}{\longrightarrow}&
4_{\rm C} ~2_{\rm L} ~1_{\rm R} ~1_{\rm V'}
&\stackrel{}{\longrightarrow}& 
\mbox{{\rm Eq.~(\ref{eq:ps6}d)}}\\
&\stackrel{1}{\longrightarrow}& 3_{\rm C} ~2_{\rm L}
~2_{\rm R} ~1_{\rm B-L} ~1_{\rm V'} &\stackrel{}{\longrightarrow}&
\mbox{{\rm Eq.~(\ref{eq:ps6}c)}}\\
&\stackrel{1}{\longrightarrow}& 3_{\rm C} ~2_{\rm L} ~1_{\rm R} ~1_{\rm
B-L} ~1_{\rm V'} &\stackrel{}{\longrightarrow}& 
\mbox{{\rm Eq.~(\ref{eq:ps6}b)}}\\
&\stackrel{1}{\longrightarrow}& 3_{\rm C} ~2_{\rm L} ~1_{\rm R} ~1_{\rm
B-L} &\stackrel{2}{\longrightarrow}& {\rm G}_{\rm SM}~Z_2\\
&\stackrel{1}{\longrightarrow}& {\rm G}_{\rm SM} ~1_{\rm V}  &\stackrel{2}{\longrightarrow}& {\rm G}_{\rm SM}~Z_2 \\
&\stackrel{1,2}{\longrightarrow}& {\rm G}_{\rm SM} ~1_{\rm V'} ~Z_2 &\stackrel{2}{\longrightarrow}& {\rm G}_{\rm SM}~Z_2 \\
  \end{array}
  \right.
    }
\end{equation}

The SSB schemes of SU(6) and SU(7) down to the
G$_{\rm SM}$ which could accommodate an inflationary era with no
defect (of any kind) at later times are inconsistent with proton
lifetime measurements and minimal SU(6) and SU(7) do not predict
neutrino masses \cite{jrs}, implying that these models are
incompatible with high energy physics phenomenology. Higher
rank groups, namely SO(14), SU(8) and SU(9), should in general lead to
cosmic string formation at the end of hybrid inflation. In all these
schemes, cosmic string formation is sometimes accompanied by the
formation of embedded strings. The strings which form at the end of
hybrid inflation have a mass which is proportional to the inflationary
scale.

\section{Braneworld Cosmology}
\label{sec:bc}
One of our dreams in theoretical physics is to be able to unify all
fundamental interactions into a unique theory. String theory
offers one such attempt to unify gravity with the other interactions,
in a self-consistent quantum theory. String theory is based on
the proposal that one-dimensional extended objects (strings) are the
fundamental constituents of matter.  In the mid 1990's it was realised
that higher dimensional extended membranes ($p$-{\sl branes}, with
$p>1$) should also play a crucial r\^ole in string theory. In
particular, branes offer the possibility of relating apparently
different string theories.  Of particular importance among $p$-branes
are the $Dp$-branes on which open strings can end; they can describe
matter fields living on the brane. Closed strings ({\sl eg.}
graviton) live on the higher dimensional bulk; their excitations
describe perturbtions on the bulk geometry. Classically, matter
and radiation fields are localised on the brane, with gravity
propagating in the bulk.

Some of the extra dimensions could be far larger than what had been
previously thought. If the extra dimensions were testable only
via gravity then they might be relatively large, leading to a possible
explanation for the weakness of gravity as compared to the other
fundamental interactions. It has been proposed that the gravitational
field of an object could leak out into the large but hidden extra
dimensions, leading to a weaker gravity as perceived from an observer
living in a four-dimensional universe.  More precisely, the effective
value of Newton's constant in a four-dimensional universe, $G_{(4)}$,
can be written as $G_{(4)}\equiv G_{({\rm D})}/R^{D-4}$, where $D$
denotes the total dimensionality of spacetime and $R$ stands for the
radius of compactification (assumed, without loss of generality, to be
the same in all extra dimensions). The absence of any observed
deviation from the familiar Newton's law (in a four-dimensional
spacetime) imposes an upper limit on the compactification radius. More
precisely, the present experimental constraints yield $R\lsim 0.2 {\rm
mm}$.

\subsection{Inflation within Braneworld Cosmologies}
In the context of braneworld cosmology, brane
inflation occurs in a similar way as hybrid inflation within
supergravity, leading to string-like objects.  In string
theories, D-brane $\bar{\rm D}$-anti-brane annihilation leads
generically to the production of lower dimensional D-branes, with D3-
and D1-branes (D-strings) being predominant \cite{rmm}.

To sketch brane inflation (for example see Ref.~\cite{tye}), consider a
D$p$-${\bar{\rm D}}p$ system in the context of IIB string theory. Six
of the spatial dimensions are compactified on a torus; all branes move
relatively to each other in some directions. A simple and
well-motivated inflationary model is brane inflation where the
inflaton is simply the position of a D$p$-brane moving in the bulk.
As two branes approach, the open string modes between the branes
develop a tachyon, indicating an instability.  The relative
D$p$-${\bar{\rm D}}p$-brane position is the inflaton field and the
inflaton potential comes from their tensions and interactions.  Brane
inflation ends by a phase transition mediated by open string
tachyons. The annihilation of the branes releases the brane tension
energy that heats up the universe so that the hot big ban epoch can
take place. Since the tachyonic vacuum has a non-trivial $\pi_1$
homotopy group, there exist stable tachyonic string solutions with
$(p-2)$ co-dimensions. These daughter branes have all dimensions
compact; a four-dimensional observer perceives them as
one-dimensional objects, the D-strings.  Zero-dimensional defects
(monopoles) and two-dimensional ones (domain walls), which are
cosmologically undesirable, are not produced during brane
intersections.  

\subsection{Cosmic Superstrings}
The first to consider cosmic superstrings as playing the r\^ole of
cosmic strings was Witten \cite{witten}. However, since for
fundamental strings the linear mass density is proportional to (string
energy scale)$^2$, it was realised that for a string energy scale of
the order of the Planck mass, $G\mu$ becomes of the order of 1, and
therefore this proposal was ruled out since observational data require
$G\mu\lsim 10^{-7}$.  More recently, in the framework of braneworld
scenarios the large compact dimensions and the large warp factors
allow the string energy scale to be much smaller than the Planck
scale. Thus, in models with large extra dimensions, cosmic superstring
tensions could have values in the range between $10^{-13}< G\mu <
10^{-6}$, depending on the model. These cosmic suprestrings are
stable, or at least their lifetime is comparable to the age of the
universe, so they can survive to form a cosmic superstring network.

Type IIB string theory, after compactification to 3+1 dimensions, has
a spectrum of one-dimensional objects, the Fundamental (F) strings,
carrying charge under the Neveu Schwartz -- Neveu Schwartz two-form
potential, and the Dirichlet (D) strings carrying charge under the
Ramond--Ramond two-form potential.  Both these strings are
individually $\frac{1}{2}$-BPS (Bogomol'nyi-Prasad-Sommerfield)
objects, with however each type breaking a different half of the
supersymmetry.  F- and D-strings that survive the cosmological
evolution become cosmic superstrings with interesting cosmological
implications \cite{polchinski}.  Thus, string theory offers two
distinct candidates for playing the r\^ole of cosmic strings.

IIB string theory allows the existence of bound $(p,q)$ states of $p$
F-strings and $q$ D-strings, where $p$ and $q$ are coprime.  A
$(p,q)$ state is still a $\frac{1}{2}$-BPS object with tension
\begin{equation}
\mu_{(p,q)}=\mu_{\rm F}\sqrt{p^2+q^2/g_{\rm s}^2}~,
\end{equation}
where $\mu_{\rm F}$ denotes the effective F-string tension after
compactification and $g_{\rm s}$ stands for the string coupling.

Cosmic superstrings share a number of properties with cosmic strings,
but there are also differences which may lead to distinctive
observational signatures. In general, string intersections lead to
intercommutation and loop production. For cosmic strings the
probability of intercommutation ${\cal P}$ is equal to 1, whereas this
is not the case for F- and D-strings. Clearly, D-strings can miss each
other in the compact dimension, leading to a smaller ${\cal P}$, while
for F-strings the scattering has to be calculated quantum mechanically
since these are quantum mechanical objects.  The collisions between
all possible pairs of superstrings have been studied in string
perturbation theory \cite{jjp}. For F-strings, the reconnection
probability is of the order of $g_{\rm s}^2$, where $g_{\rm s}$ stands
for the string coupling.  For F-F string collisions, it was found
\cite{jjp} that the reconnection probability $\cal P$ is $10^{-3}\lsim
{\cal P}\lsim 1$. For D-D string collisions, one has
$10^{-1}\lsim{\cal P}\lsim 1$. Finally, for F-D string collisions, the
reconnection probability can take any value between 0 and 1.  These
results have been confirmed \cite{hh1} by a quantum calculation of the
reconnection probability for colliding D-strings.  Similarly, the
string self-intersection probability is reduced.

In contrast to the networks formed from Abelian strings, which consist
of loops and long strings, $(p,q)$ networks can also contain links
which start and end at a three-point vertex.  More precisely, when F-
and D-strings meet they can form a three-string junction, with a
composite FD-string.  Such links could potentially lead to a frozen
network, which could dominate the matter content of the universe.

Modelling the evolution of a $(p,q)$ network is a challenging task, in
particular due to the existence of the junctions. Nevertheless,
various attempts have been undertaken and they all conclude
\cite{ms-ep-tww} that the network will reach a {\sl scaling} regime,
in which the length scales increase in proportion to time.

Cosmic superstrings interact with the standard model particles 
via gravity, implying that their detection involves gravitational
interactions. Since the particular brane inflationary scenario remains
unknown, the tensions of superstrings are only loosely constrained.

\section{Observational Consequences}
\label{sec:oc}

\subsection{CMB Temperature Anisotropies}
\label{subsec:cmb}
The CMB temperature anisotropies offer a powerful test for theoretical
models aiming at describing the early universe.  The characteristics
of the CMB multipole moments can be used to discriminate among
theoretical models and to constrain the parameters space.

The spherical harmonic expansion of the CMB temperature anisotropies,
as a function of angular position, is given by
\begin{equation}
\label{dTT}
\frac{\delta T}{T}({\bf n})=\sum _{\ell m}a_{\ell m} {\cal W}_\ell
Y_{\ell m}({\bf n})~\,
\ \ \ \mbox {with}\ \ \ 
a_{\ell m}=\int {\rm
d}\Omega _{{\bf n}}\frac{\delta T}{T}({\bf n})Y_{\ell m}^*({\bf n})~;
\end{equation}
${\cal W}_\ell $ stands for the $\ell$-dependent window function of
the particular experiment.  The angular power spectrum of CMB
temperature anisotropies is expressed in terms of the dimensionless
coefficients $C_\ell$, which appear in the expansion of the angular
correlation function in terms of the Legendre polynomials $P_\ell$:
\begin{equation}
\biggl \langle 0\biggl |\frac{\delta T}{T}({\bf n})\frac{\delta T}{
T}({\bf n}') \biggr |0\biggr\rangle \left|_{{~}_{\!\!({\bf n\cdot
n}'=\cos\vartheta)}}\right. = \frac{1}{4\pi}\sum_\ell(2\ell+1)C_\ell
P_\ell(\cos\vartheta) {\cal W}_\ell^2 ~.
\label{dtovertvs}
\end{equation}
It compares points in the sky separated by an angle $\vartheta$.  In
Eq.~(\ref{dTT}) the brackets denote spatial average, or expectation
values if perturbations are quantised. Equation (\ref{dtovertvs})
holds only if the initial state for cosmological perturbations of
quantum-mechanical origin is the vacuum \cite{jrsgms}.  The value of
$C_\ell$ is determined by fluctuations on angular scales of the order
of $\pi/\ell$. The angular power spectrum of anisotropies observed
today is usually given by the power per logarithmic interval in
$\ell$, plotting $\ell(\ell+1)C_\ell$ versus $\ell$.

On large angular scales, the main contribution to the CMB temperature
anisotropies is given by the Sachs-Wolfe effect. Thus,
\begin{equation}
\label{sw}
\frac{\delta T}{T}({\bf n})\simeq
\frac{1}{3}\Phi [\eta _{\rm lss},{\bf n}(\eta _0-\eta _{\rm lss})]~;
\end{equation}
$\Phi (\eta ,{\bf x})$ denotes the Bardeen potential, $\eta _0$ and
$\eta _{\rm lss}$ stand for the conformal time at present and at the last
scattering surface, respectively. 

Studies of the characteristics of the CMB spectrum (amplitude and
position of acoustic peaks), in the framework of topological defect models,
have been performed even before receiving any data. Let me discuss
briefly the differences such models have, as compared to the adiabatic
perturbations induced from the amplification of the quantum
fluctuations of the inflaton field at the end of inflation, and the
difficulties one faces to extract the predictions.

For models with topological defects, perturbations are generated by
{\sl seeds} (sources), defined as any non-uniformly distributed form
of energy, which contributes only a small fraction to the total energy
density of the universe and which interacts with the cosmic fluid only
gravitationally.  Such models lead to isocurvature density
perturbations, in the sense that the total density perturbation
vanishes, but those of the individual particle species do not.
Moreover, in models with topological defects,
fluctuations are generated continuously and evolve according to
inhomogeneous linear perturbation equations.  

The energy momentum tensor of defects is determined by the their
evolution which, in general, is a non-linear process. These
perturbations are called {\sl active} and {\sl incoherent}. Active
since new fluid perturbations are induced continuously due to the
presence of the defects; incoherent since the randomness of the
non-linear seed evolution which sources the perturbations can destroy
the coherence of fluctuations in the cosmic fluid.  The highly
non-linear structure of the topological defect dynamics makes the
study of the evolution of these causal (there are no correlations on
super-horizon scales) and incoherent initial perturbations much more
complicated.

 Within linear cosmological perturbation theory, structure
formation induced by seeds is determined by the solution of the
inhomogeneous equation
\begin{equation} 
{\cal D} X({\bf k},t) = {\cal S}({\bf k},t)~, 
\end{equation}
where $X$ is a vector containing all the background perturbation
variables for a given mode specified by the wave-vector ${\bf k}$,
like the $a_{lm}$'s of the CMB anisotropies, the dark matter density
fluctuation, the peculiar velocity potential etc., ${\cal D}$ is a linear
time-dependent ordinary differential operator, and the source term 
${\cal S}$ is given by linear combinations of the energy momentum tensor of
the seed (the type of topological defects we are considering).  The
generic solution of this equation is given in terms of a Green's
function and has the following form \cite{vest}
\begin{equation}
X_i({\bf k},t_0) = \int_{t_{in}}^{t_0}{\cal G}_{il}({\bf k},t_{0},t)
{\cal S}_l({\bf k}, t){\rm d}t~.  
\end{equation} 
At the end, we need to determine expectation values, which are given
by
\begin{equation}  
\langle X_i({\bf k} ,t_0)X_j({\bf k},t_0)^*\rangle
=\int_{t_{in}}^{t_0}\int_{\eta_{in}}^{\eta_0}
{\cal G}_{il}(t_{0},t){\cal G}_{jm}^*(t_{0},t') \langle
{\cal S}_l(t){\cal S}_m^*(t')\rangle {\rm d}t {\rm d}t' .
\label{pow}
\end{equation} 
Thus, the only information we need from topological defects
simulations in order to determine cosmic microwave background and
large-scale structure power spectra, is the {\sl unequal time
two-point correlators} \cite{cor}, $\langle {\cal S}_l(t){\cal
S}_m^*(t')\rangle$, of the seed energy-momentum tensor. This problem
can, in general, be solved by an eigenvector expansion method
\cite{pen}.

On large angular scales ($\ell \leq 50$), defect models lead to the
same prediction as inflation, namely, they both predict an
approximately scale-invariant (Harrison-Zel'dovich) spectrum of
perturbations. Their only difference concerns the statistics of the
induced fluctuations. Inflation predicts generically Gaussian
fluctuations, whereas in the case of topological defect models, even
if initially the defect energy-momentum tensor would be Gaussian,
non-Gaussianities will be induced from the non-linear defect
evolution. Thus, in defect scenarios, the induced fluctuations are
non-Gaussian, at least at sufficiently high angular resolution.  This
is an interesting fingerprint, even though difficult to test through
the data.

On intermediate and small angular scales however, the predictions
of models with seeds are quite different than those of inflation, due
to the different nature of the induced perturbations.  In topological
defect models, defect fluctuations are constantly generated by the seed
evolution.  The non-linear defect evolution and the fact that the
random initial conditions of the source term in the perturbation
equations of a given scale leak into other scales, destroy perfect
coherence.  The incoherent aspect of active perturbations does not
influence the position of the acoustic peaks, but it does affect the
structure of secondary oscillations, namely secondary oscillations may
get washed out. Thus, in topological defect models, incoherent
fluctuations lead to a single bump at smaller angular scales (larger
$\ell$), than those predicted within any inflationary scenario.  This
incoherent feature is shared in common by local and global defects.

Let me briefly summarise the results: Global ${\cal O}(4)$ textures
lead to a position of the first acoustic peak at $\ell\simeq 350$ with
an amplitude $\sim 1.5$ times higher than the Sachs-Wolfe plateau
\cite{rm}.  Global ${\cal O}(N)$ textures in the large $N$ limit lead
to a quite flat spectrum, with a slow decay after $\ell \sim 100$
\cite{dkm}. Similar are the predictions of other global ${\cal O}(N)$
defects \cite{num}. (For a general study of the CMB anisotropies form
scaling seed perturbations the reader is referred to
Ref.~\cite{ruthmairi}). Local cosmic strings lead to a a power spectrum
with a roughly constant slope at low multipoles, rising up to a single
peak, with subsequent decay at small scales \cite{mark}.

At this point, I would like to bring to the attention of the reader
that the B-mode of the polarisation spectrum may be a smoking gun for
the cosmic strings \cite{mark}, since inflation gives just a weak
contribution. The reason being that scalar modes may contribute to the
B-mode only through the gravitational lensing of the E-mode. Thus, the
large vector contribution from cosmic strings may lead in the future
to the detection of strings.

The position and amplitude of the acoustic peaks, as found by the CMB
measurements (see, e.g. Ref.~\cite{wmap3}), are clearly in
disagreement with the predictions of topological defect models. Thus,
CMB measurements rule out pure topological defect models as the unique
origin of initial density perturbations leading to the observed
structure formation. However, since strings and string-like defects
are generically formed, then one should consider them as a
sub-dominant partner of inflation. Thus, one should study the
compatibility between {\sl mixed} perturbation models \cite{bprs} and
observational data.

Consider  therefore a model in which a network of cosmic
strings evolved independently of any pre-existing fluctuation
background, generated by a standard cold dark matter with a non-zero
cosmological constant ($\Lambda$CDM) inflationary phase. Restrict 
your attention to the angular spectrum, so that you are in the
linear regime.  Thus,
\begin{equation}
C_\ell =   \alpha     C^{\scriptscriptstyle{\rm I}}_\ell
         + (1-\alpha) C^{\scriptscriptstyle{\rm S}}_\ell~,
\label{cl}
\end{equation}
where $C^{\scriptscriptstyle{\rm I}}_\ell$ and $C^{\scriptscriptstyle
{\rm S}}_\ell$ denote the (COBE normalised) Legendre coefficients due
to adiabatic inflaton fluctuations and those stemming from the string
network, respectively. The coefficient $\alpha$ in Eq.~(\ref{cl}) is a
free parameter giving the relative amplitude for the two
contributions.  Then one has to compare the $C_\ell$, given by
Eq.~(\ref{cl}), with data obtained from CMB anisotropy
measurements. The inflaton and string induced uncorrelated spectra as
a function of $\ell$ , both normalised on the COBE data, together with
the weighted sum, are shown in Fig.~(\ref{fig:fig5}) (see,
Ref.~\cite{bprs}).
\vskip 0.5cm
\begin{figure}
\centering \includegraphics[width=2.4in,angle=270]{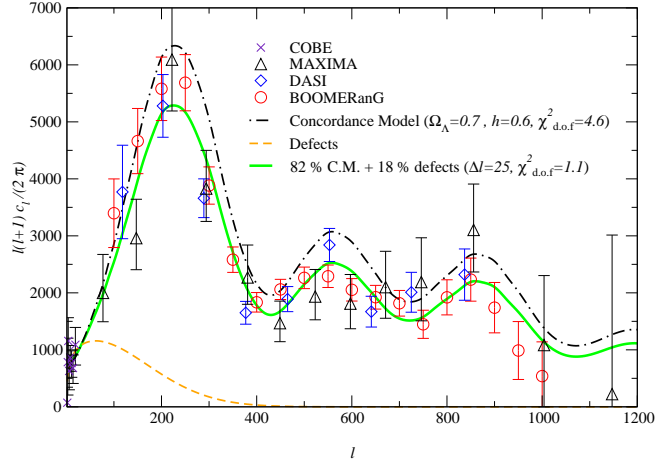}
\vskip.4cm\caption{$\ell (\ell + 1) C_\ell$ versus $\ell$ for three
    different models. The upper dot-dashed line represents the
    prediction of a $\Lambda$CDM model. The lower dashed line is a
    typical string spectrum. Combining both curves with the
    extra-parameter $\alpha$ produces the solid curve, with a $\chi^2$
    per degree of freedom slightly above unity. The string
    contribution turns out to be some $18\%$ of the total
    \cite{bprs}.}
\label{fig:fig5}
\end{figure} 
The quadrupole anisotropy due to {\sl freezing in} of quantum
fluctuations of a scalar field during inflation reads
\begin{equation}\label{contribInfl}
\left(\frac{\delta T}{T}\right)_{\rm Q-infl} =
\left[\left(\frac{\delta T}{T}\right)_{\rm Q- scal}^2 +
\left(\frac{\delta T}{T}\right)_{\rm Q- tens}^2\right]^{1/2}~,
\end{equation}
with the scalar and tensor contributions given by
\begin{equation}\label{contribInflScal}
\left(\frac{\delta T}{T}\right)_{\rm Q- scal} =
\frac{1}{4\sqrt{45}\pi}\frac{V^{3/2}(\varphi_Q)}{M_{\rm
Pl}^3\,V'(\varphi_Q)}~,
\end{equation}
and 
\begin{equation}
\label{contribInflTens}
\left(\frac{\delta T}{T}\right)_{\rm Q-tens}\sim {0.77\over 8\pi}
\,\frac{V^{1/2}(\varphi_Q)}{M_{\rm Pl}^2}~,
\end{equation}
respectively.  Here $V$ is the potential of the inflaton field
$\varphi$, with $V'\equiv {\rm d}V(\varphi)/{\rm d}\varphi$, 
$M_{\rm Pl}$ denotes the reduced Planck mass, $M_{\rm Pl}= {(8\pi
G)^{-1/2}}\simeq 2.43\times 10^{18}$ GeV, and $\varphi_{\rm Q}$ is the
value of the inflaton field when the comoving scale corresponding to
the quadrupole anisotropy became bigger than the Hubble radius.

Simulations of Goto-Nambu local strings in a
Friedmann--Lema\^{\i}tre--\-Ro\-berston--Walker spacetime lead to
\cite{ls2003}
\begin{equation}\label{contribCS}
\left(\frac{\delta T}{T}\right)_{\rm cs}\sim (9-10) G\mu \quad
\mathrm{with}\quad \mu=2\pi\langle\chi \rangle^2~,
\end{equation}
where $\langle\chi \rangle$ is the vacuum expectation value of the
Higgs field responsible for the formation of cosmic strings.

Before discussing F- and D-term inflation, I would like briefly to
describe the {\sl curvaton} mechanism \cite{lw2002}, according which
the primordial fluctuations could also be generated from the quantum
fluctuations of a late-decaying scalar field, the {\sl curvaton} field
${\cal \psi}$, which does not play the r\^ole of the inflaton field.
During inflation the curvaton potential is very flat and the curvaton
acquires quantum fluctuations, which are expressed in terms of the the
expansion rate during inflation, $H_{\rm infl}=\sqrt(8\pi G/ 3)
V(\varphi)$, through
\begin{equation}
\delta{\cal \psi}_{\rm init}={H_{\rm inf}\over 2\pi}~.
\label{qfc}
\end{equation}
They lead to entropy fluctuations at the end of inflation.

During the radiation-dominated era the curvaton decays and reheats the
universe. The primordial fluctuations of the curvaton field are
converted to purely adiabatic density fluctuations, thus
the curvaton contribution in terms of the metric perturbation reads
\begin{equation}
\left(\frac{\delta T}{T}\right)_{\rm curv}
=-{4\over 27}{\delta{\cal \psi_{\rm init}}\over
\psi_{\rm init}}~.
\label{Psi_curv}
\end{equation}
If one assumes the additional contribution to the temperature
anisotropies originated from the curvaton field, then
\begin{equation}
\label{totalcmb}
\left[\left(\frac{\delta T}{T}\right)_{\rm tot}\right]^2 =
\left[\left(\frac{\delta T}{T}\right)_{\rm infl}\right]^2 +
\left[\left(\frac{\delta T}{T}\right)_{\rm cs}\right]^2+
\left[\left(\frac{\delta T}{T}\right)_{\rm curv}\right]^2
~.
\end{equation}
The total quadrupole anisotropy, the {\sl l.h.s.} of
Eq.~(\ref{totalcmb}), is the one to be normalised to the Cosmic
Background Explore (COBE) data \cite{cobe}, namely $\left(\delta T/
T\right)_{\rm Q}^{\rm COBE} \sim 6.3\times 10^{-6}$.

\subsubsection{F-term Inflation}
Considering only large angular scales one can calculate the contributions
to the CMB temperature anisotropies analytically.  The quadrupole
anisotropy has one contribution coming from the inflaton field,
calculated using Eq.~(\ref{VexactF}), and one contribution coming from
the cosmic string network.  Fixing the number of e-foldings to 60, 
the inflaton and cosmic string contributions to the CMB depend on the
superpotential coupling $\kappa$, or equivalently, on the symmetry
breaking scale $M$ associated with the inflaton mass scale, which
coincides with the string mass scale. 

The total quadrupole anisotropy, to be normalised to the COBE data, is
found to be \cite{rs1}
\begin{eqnarray}
\label{eqnumF}
\left({\delta T\over T}\right)_{\rm Q-tot} &\sim&\Big\{y_{\rm
Q}^{-4}\left({\kappa^2 \mathcal{N}\, N_{\rm Q}\over 32\pi^2}\right)^2
\Big[\frac{64N_{\rm Q}}{45\cal N} x_{\rm Q}^{-2}y_{\rm
Q}^{-2}f^{-2}(x_Q^2)\nonumber\\
&&~~~~~~~~~~~~~~~~~~~~~~~~~+\left(\frac{0.77 \kappa}{\pi}\right)^2 +
324\Big]\Big\}^{1/2}~.
\end{eqnarray}
In Eq.~(\ref{eqnumF}),
\begin{equation}
x_{\rm Q}={|S_{\rm Q}|\over M}~~;~~ y_{\rm
Q}^2=\int_1^{x_{\rm Q}^2}\frac{{\rm d}z}{zf(z)}
\end{equation}
and
\begin{equation}
 N_{\rm Q}=\frac{4\pi^2}{\kappa^2\cal N}\frac{M^2}{M_{\rm
Pl}^2}\,y_{\rm Q}^2~,
\end{equation}
with
\begin{equation}
f(z)=(z+1)\ln(1+z^{-1})+(z-1)\ln(1-z^{-1})~.
\end{equation}
As noted earlier, the index $_{\rm Q}$ denotes the scale responsible
for the quadrupole anisotropy in the CMB.

The cosmic string contribution is consistent with the CMB measurements
provided \cite{rs1}
\begin{equation}
\label{kappaFCMB}
M\lsim 2\times 10^{15} {\rm GeV} ~~\Leftrightarrow ~~\kappa \lsim
7\times10^{-7}~.
\end{equation}
Strictly speaking the above condition was found in the context of
SO(10) gauge group, but the conditions imposed in the case of other
gauge groups are of the same order of magnitude since $M$ is a slowly
varying function of the dimensionality ${\cal N}$ of the
representations to which the scalar components of the chiral Higgs
superfields belong \cite{rs1}.

The superpotential coupling $\kappa$ is also subject to the gravitino
constraint, which imposes an upper limit to the reheating temperature
to avoid gravitino overproduction. Within the framework of SUSY GUTs
and assuming the see-saw mechanism to give rise to massive neutrinos,
the inflaton field decays during reheating into pairs of right-handed
neutrinos.  This constraint on the reheating temperature can be
converted into a constraint on the superpotential coupling
$\kappa$. The gravitino constraint on $\kappa$ reads \cite{rs1}
$\kappa \lsim 8\times 10^{-3}$, which is a weaker constraint than the
one obtained from the CMB, Eq.~(\ref{kappaFCMB}).

The tuning of the free parameter $\kappa$ can be softened if one
allows for the curvaton mechanism.  Clearly, within supersymmetric
theories such scalar fields are expected to exist. In addition,
embedded strings, if they accompany the formation of cosmic strings,
they may offer a natural curvaton candidate, provided the decay
product of embedded strings gives rise to a scalar field before the
onset of inflation.  Considering the curvaton scenario, the coupling
$\kappa$ is only constrained by the gravitino limit. More precisely,
assuming the existence of a curvaton field there is an additional
contribution to the temperature anisotropies.  Calculating the
curvaton contribution to the temperature anisotropies, one obtains
the additional contribution \cite{rs1}
\begin{equation}\label{curvterm}
\left[\left({\delta T\over T}\right)_{\rm curv}\right]^2 =
y_Q^{-4}\,\left({\kappa^2 \mathcal{N} N_{\rm Q}\over 32\pi^2} \right)^2\, 
\left[ \left(\frac{16}{81\pi\sqrt3}\right)\,
\kappa\, \left(\frac{M_{\rm Pl}}{{\cal\psi}_{\rm init}}\right)\right]^2~.
\end{equation}
Normalising the total $\left(\delta T/ T\right)_{\rm Q}$ ({\sl i.e.}
the inflaton, cosmic string and curvaton contributions) to the data
one gets \cite{rs1} the following limit on the initial value of the
curvaton field
\begin{equation}
{\cal\psi}_{\rm init} \lsim 5\times 10^{13}\,\left( 
\frac{\kappa}{10^{-2}}\right){\rm GeV}~~\mbox{for}~~
 \kappa\in [10^{-6},~1].
\end{equation}
Finally, I would like to point out that in the case of F-term
inflation\footnote {This does not hold for D-term inflation; the
strings formed at the end of D-term inflation are BPS-objects.}, the
linear mass density $\mu$ (see, Eq.~(\ref{contribCS})) gets a
correction due to deviations from the Bogomol'nyi limit, enlarging the
parameter space for F-term inflation \cite{enlarge}.  More precisely,
this correction to $\mu$ turns out to be proportional to
$\ln(2/\beta)^{-1}$, where $\beta$ is proportional to the square of
the ratio between the superpotential and the GUT couplings. Thus,
under the assumption that strings contribute less than 10$\%$ to the
power spectrum at $\ell=4$, the bound on $\kappa$ reduces to the one
imposed by the gravitino limit.

\subsubsection{D-term Inflation}
D-term inflation leads to cosmic string formation at the end of the
inflationary era.  The total quadrupole temperature anisotropy, to be
normalised to the COBE data, reads \cite{rs1}
\begin{eqnarray}\label{eqnumDsugra}
\left({\delta T\over T}\right)_{\rm Q}^{\rm tot} &\sim&
\frac{\xi}{M_{\rm Pl}^2}\Big\{ \frac{\pi^2}{90g^2}x^{-4}_{\rm
Q}f^{-2}(x^2_{\rm Q}) \frac{{\rm W}(x^2_{\rm Q}(g^2 \xi)(\lambda^2 M_{\rm
Pl}^2))} {\left[1+{\rm W}(x^2_{\rm Q}(g^2 \xi)(\lambda^2 M_{\rm
Pl}^2))\right]^2}\nonumber\\
&&~~~~~~~
 +\left(\frac{0.77
g}{8\sqrt{2}\pi}\right)^2 +
\left(\frac{9}{4}\right)^2\Big\}^{1/2}~,
\end{eqnarray}
where the only unknown is the Fayet-Iliopoulos term $\xi$, for given
values of $g$ and $\lambda$. Note that ${\rm W}(x)$ is the ``W-Lambert
function'', i.e. the inverse of the function $F(x)=xe^x$. Thus, one
can get $\xi$ numerically, and then obtain $x_{\rm Q}$, as well as the
inflaton and cosmic string contribution, as a function of the
superpotential and gauge couplings $g$ and $\lambda$.
In the case of minimal SUGRA, consistency between
CMB measurements and theoretical predictions impose \cite{rs1,prl2005}
\begin{equation}
g\lsim 2\times 10^{-2}~ ~ \mbox{and}~ ~ \lambda\lsim 3\times
10^{-5}~,
\end{equation} 
which can be expressed as a single constraint on the Fayet-Iliopoulos
term $\xi$,
\begin{equation}
\sqrt\xi \lsim 2\times 10^{15}~{\rm GeV}~.
\end{equation}
These results are shown in Fig.~(\ref{fig:fig6}).
\begin{figure}
\centering
\includegraphics[height=4cm]{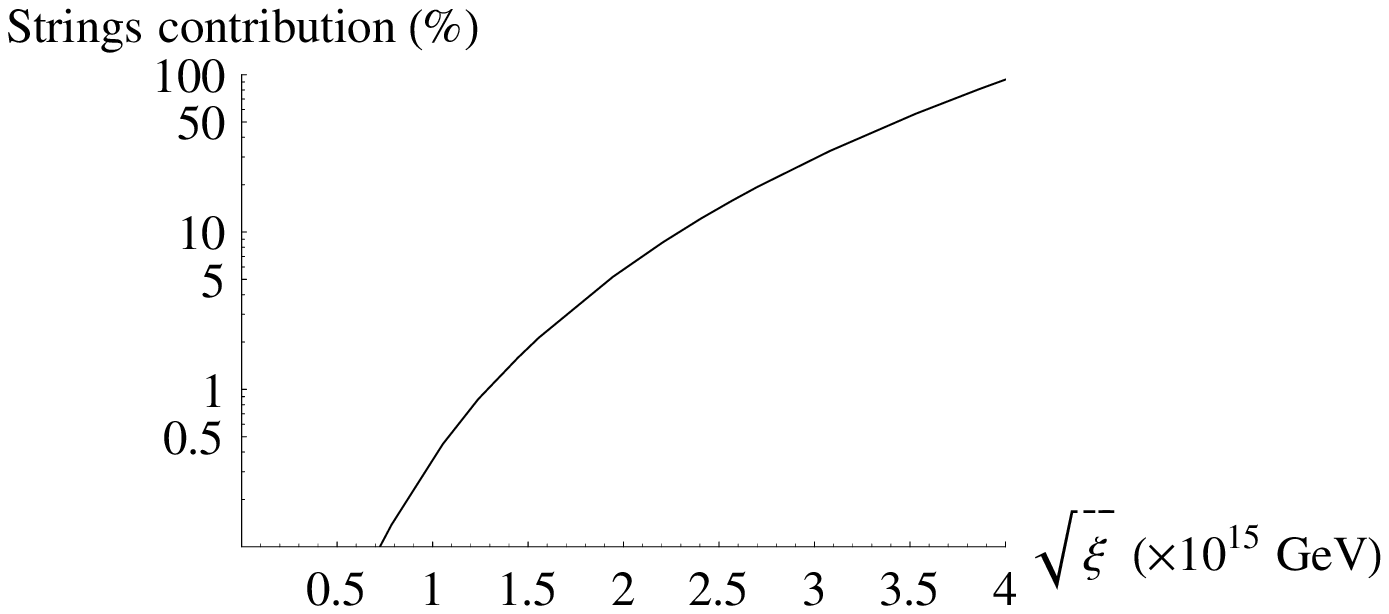}
\vskip.5cm
\includegraphics[height=4cm]{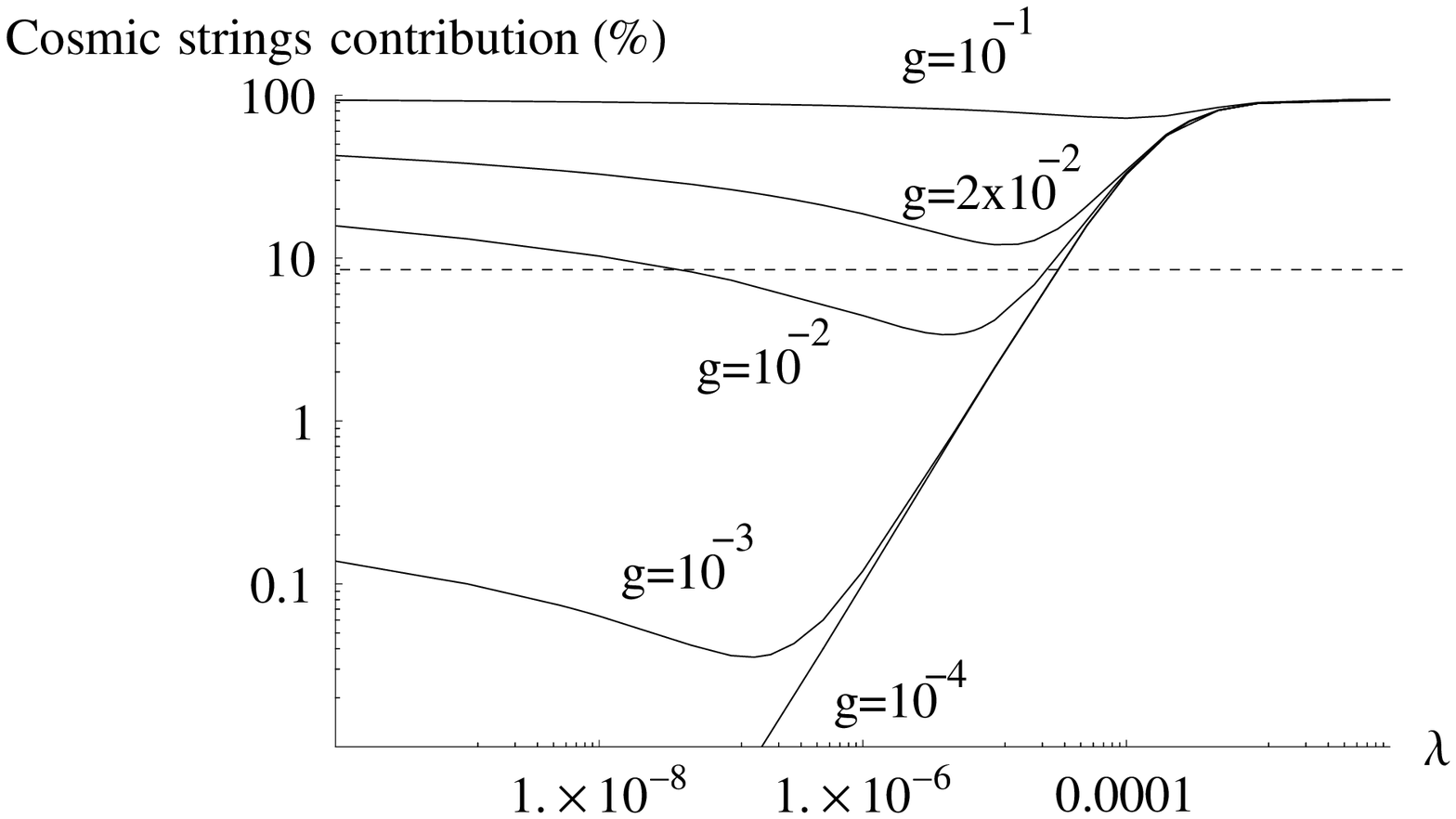}
\caption{At the top, the cosmic string contribution to he CMB, as a
function of the mass scale $\sqrt\xi$ in units of $10^{15} {\rm GeV}$.
At the bottom, cosmic string contribution to the CMB temperature
anisotropies, as a function of the superpotential coupling $\lambda$,
for different values of the gauge coupling $g$. The maximal
contribution allowed by WMAP is represented by a dotted line
\cite{prl2005}.  }
\label{fig:fig6}       
\end{figure}

The fine tuning on the couplings can be softened if one invokes the
curvaton mechanism.  Calculating the curvaton contribution to the
temperature anisotropies, one obtains the additional contribution
\cite{prl2005}
\begin{equation}\label{curvtermD}
\left[\left({\delta T\over T}\right)_{\rm curv}\right]^2 =
\frac{1}{6}\,
 \left(\frac{2}{27\pi}\right)^2\,
 \left(\frac{g\xi}{M_{\rm Pl}\psi_{\rm init}}\right)^2~.
\end{equation}
Thus, the gauge coupling can reach the upper bound imposed from the
gravitino mechanism, provided the initial value of the curvaton field
is \cite{prl2005}
\begin{equation}
\psi_{\rm init}\lsim 3\times 10^{14}\left(\frac{g}{10^{-2}}\right){\rm
GeV}~~\mbox{for}~~\lambda\in[10^{-1},10^{-4}]~;
\end{equation}
for smaller values of $\lambda$, the curvaton mechanism is not
necessary.  This result is explicitly shown in Fig.~(\ref{fig:fig7}).
\begin{figure}
\centering
\includegraphics[height=4cm]{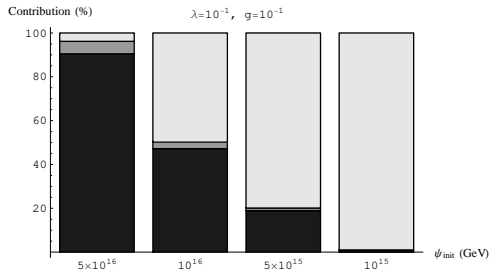}
\caption{The cosmic string (dark grey), curvaton (light
grey) and inflaton (grey) contributions to the CMB temperature
anisotropies as a function of the the initial value of the curvaton
field ${\cal\psi}_{\rm init}$, for $\lambda=10^{-1}$ and $g=10^{-1}$
\cite{prl2005}.  }\label{fig:fig7}
\end{figure} 

Concluding, within minimal supergravity the couplings and masses must
be fine tuned to achieve compatibility between measurements on the CMB
temperature anisotropies and theoretical predictions.  Note that for
minimal D-term inflation, one can neglect the corrections introduced by
the superconformal origin of supergravity.

The constraints on the couplings remain qualitatively valid in
non-minimal supergravity theories; the superpotential $W$ given in
Eq.~(\ref{superpoteninflaD}) and we consider a non-minimal K\"ahler
potential.  Let us first consider D-term inflation based on K\"ahler
geometry with shift symmetry.  If we identify the inflaton field with
the real part of $S$ then we obtain the same constraint for the
superpotential coupling as in the minimal supergravity case. However,
if the inflaton field is the imaginary part of $S$, then we get that
the the cosmic string contribution becomes dominant, in contradiction
with the CMB measurements, unless the superpotential coupling is
\cite{rs3}
\begin{equation}
\lambda\lsim 3\times 10^{-5}~.
\end{equation}
We show this constraint in Fig.~(\ref{fig:fig8}).
\begin{figure}
\centering
\includegraphics[height=4cm]{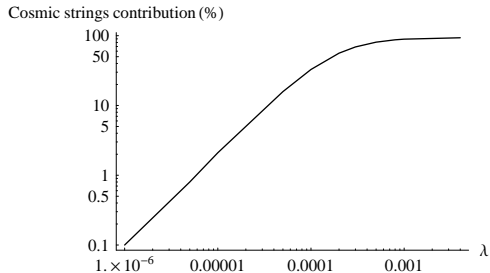}
\caption{Cosmic string contribution to the CMB temperature
anisotropies as a function of $\lambda$, in the case of D-term
inflation based on a K\"ahler geometry with shift symmetry. The
inflaton field is identified with the imaginary part \cite{rs3}.
}\label{fig:fig8}
\end{figure} 

Considering D-term inflation based on a K\"ahler potential with
non-renormalisable terms, the contribution of cosmic strings dominates
if the superpotential coupling $\lambda$ is close to unity. Setting
$f_\pm(|S|^2/M_{\rm Pl}^2)=c_\pm(|S|^2/M_{\rm Pl}^2)$, we find that in
the simplified case $b=0$ (see, Eq.~(\ref{gen})), the constraints on
$\lambda$ read \cite{rs3}
\begin{equation}\label{contraintelambdanonmin}
(0.1-5)\times 10^{-8} 
 \leq \lambda \leq
(2-5)\times 10^{-5} 
\end{equation}
or equivalently
\begin{equation}
\sqrt{\xi}\leq 2\times 10^{15}
\;\mathrm{GeV}~,
\end{equation}
implying 
 \begin{equation}
G\mu \leq 8.4\times 10^{-7}~.
\end{equation}
In the general case, where $b\neq0$, the constraints are shown in
Fig.~(\ref{fig:fig9}).
\begin{figure}
\centering
\includegraphics[height=4cm]{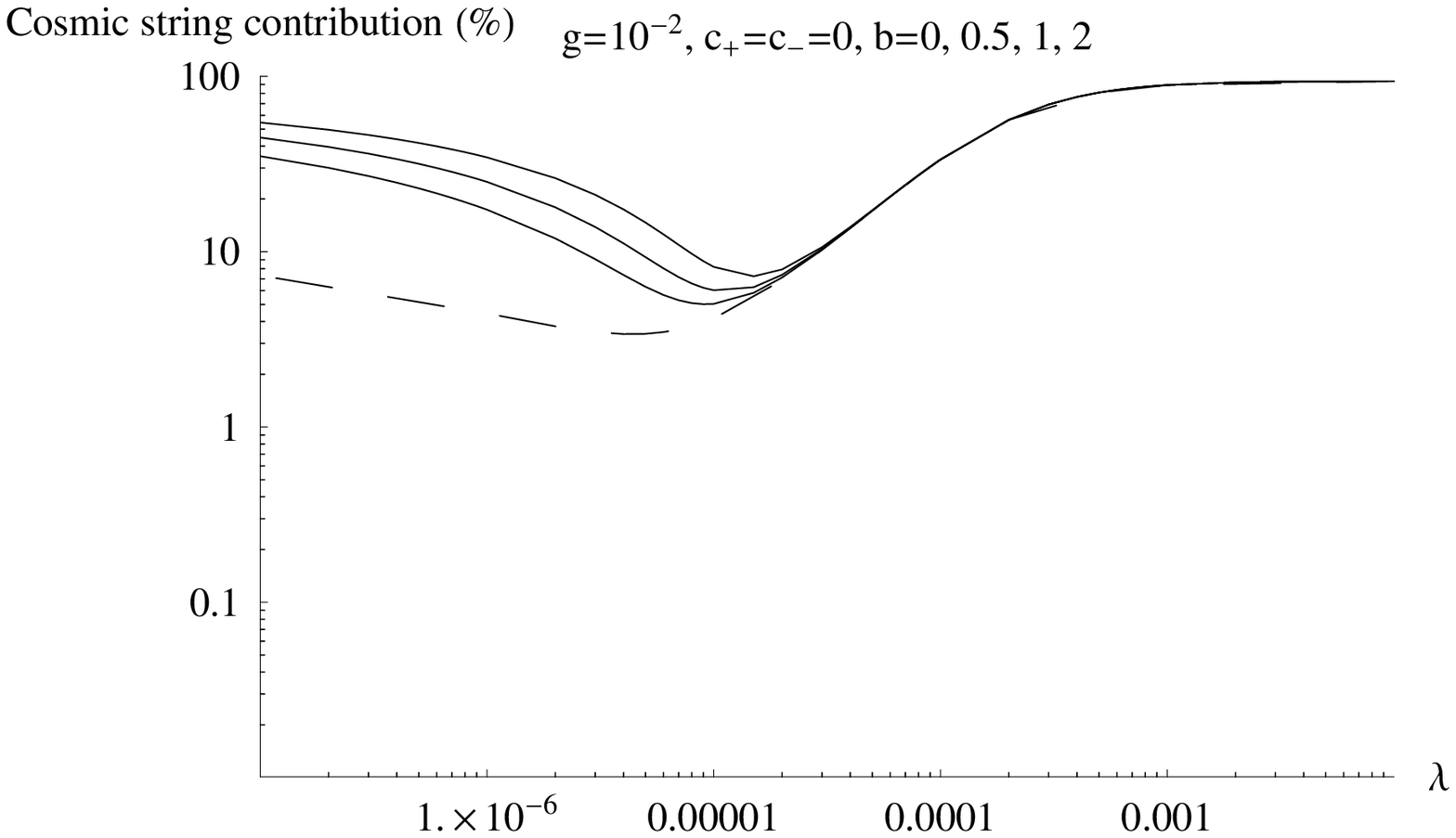}
\vskip.8cm
\includegraphics[height=4cm]{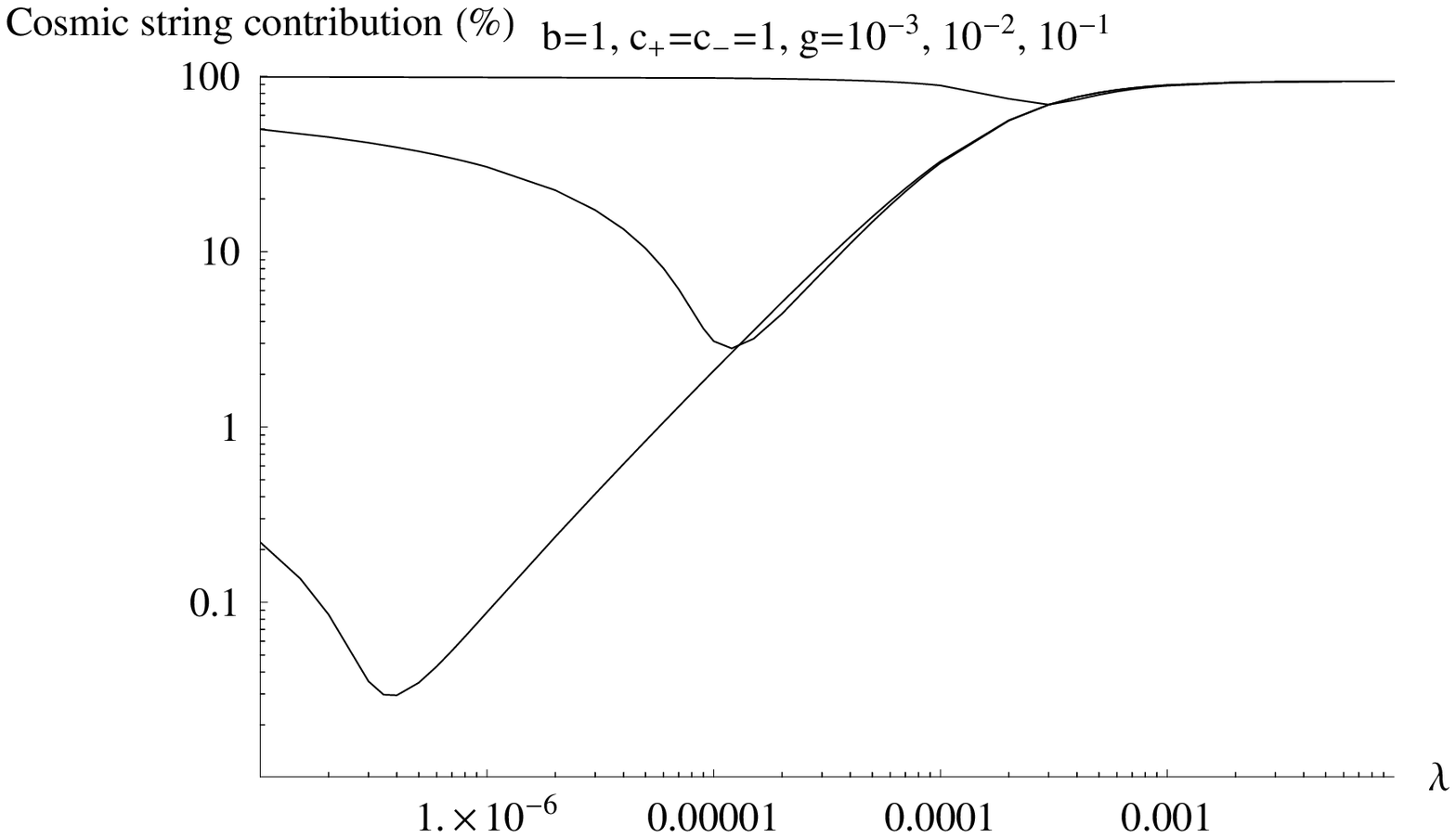}
\caption{Cosmic string contribution to the CMB temperature
anisotropies as a function of $\lambda$, in the case of D-term
inflation based on a K\"ahler potential with non-renormalisable terms
\cite{rs3}.  In the top panel we set $g=10^{-2}$ and $c_\pm =0$; the
simple case $b=0$ is represented by the dashed line, while plain lines
show the contributions for $b=0.5, 1, 2$, going from the bottom to the
top \cite{rs3}. In the bottom panel we set $b=1$ and $c_\pm=1$; the
plain lines show the contributions for $g=10^{-3}, 10^{-2}, 10^{-1}$,
going from bottom to the top \cite{rs3}.}
\label{fig:fig9}
\end{figure}

In conclusion, higher order K\"ahler potentials do {\bf not} suppress
cosmic string contribution, as it was incorrectly claimed in the
literature.  By allowing a small, but non-negligible, contribution of
strings to the angular power spectrum of CMB anisotropies, we
constrain the couplings of the inflationary models, or equivalently
the dimensionless string tension. These models remain compatible with
the most current CMB measurements, even when one calculates \cite{ns}
the spectral index. More precisely, the inclusion of a sub-dominant
string contribution to the large scale power spectrum amplitude of the
CMB, increases the preferred value for the spectral index \cite{ns}. 

\subsubsection{Brane Inflation}
The CMB temperature anisotropies originate from the amplification of
quantum fluctuations during inflation, as well as from the cosmic
superstring network. If the scaling regime of the superstring
network is the unique source of the density perturbations, the COBE
data yield $G\mu \simeq 10^{-6}$. Using the latest WMAP data, the 
contribution from strings to the  total CMB power spectrum on
observed scales is at most 10$\%$, which translates in the upper limit on the
dimensionless string tension $G\mu\lsim 1.8  \ (2.7)
\times 10^{-7}$ at $68  \ (95)\%$ confidence \cite{pog}. Thus, the cosmic
superstrings produced towards the end of inflation in the context of
braneworld cosmological models is in agreement with the present CMB
data.

\subsection{Gravitational Wave Background}

Oscillating cosmic string loops emit \cite{vil} Gravitational Waves
(GW). Long strings are not straight but they have a superimposed wiggly
small-scale structure due to string intercommutations, thus they also
emit \cite{ms-gw} GW. Cosmic superstrings can also generate~\cite{dv}
a stochastic GW background. Therefore, provided the emission of
gravity waves is the efficient mechanism \cite{cmf,gwandstrings} for
the decay of string loops, cosmic strings/superstrings could provide a
source for the stochastic GW spectrum in the low-frequency band.  The
stochastic GW spectrum has an almost flat region in the frequency
range $10^{-8}-10^{10}$ Hz. Within this window, both ADVANCED
LIGO/VIRGO (sensitive at a frequency $f\sim 10^2$ Hz) and LISA
(sensitive at $f\sim 10^{-2}$ Hz) interferometers may have a chance of
detectability.

Strongly focused beams of relatively high-frequency GW are emitted by
cusps and kinks in oscillating strings/superstrings. The distinctive
waveform of the emitted bursts of GW may be the most sensitive test of
strings/superstrings. ADVANCED LIGO/VIRGO may detect bursts of GW for
values of $G\mu$ as low as $10^{-13}$, and LISA for values down to
$G\mu\geq 10^{-15}$. At this point, I would like to remind to the
reader that there is still a number of theoretical uncertainties for
the evolution of a string/superstring network \cite{gwandstrings}.

Recently, they have been imposed limits \cite{pulsar} on an isotropic
gravitational wave background using pulsar timing observations, which
offer a chance of studying low-frequency (in the range between
$10^{-9}-10^{-7}$ Hz) gravitational waves.  The imposed limit on the
energy density of the background per unit logarithmic frequency
interval reads $\Omega_{\rm GW}^{\rm cs}(1/8{\rm yr})h^2\leq 1.9\times
10^{-8}$ (where $h$ stands for the dimensionless amplitude in GW
bursts).  

If the source of the isotropic GW background is a cosmic
string/superstring network, then it leads to an upper bound on the
dimensionless tension of a cosmic string/superstring background. Under
reasonable assumptions for the string network the upper bound on the
string tension reads \cite{pulsar} $G\mu\leq 1.5\times 10^{-8}$. This
is a strongest limit than the one imposed from the CMB temperature
anisotropies. Thus, F- and D-term inflation become even more fine
tuned, unless one invokes the curvaton mechanism.

This limit does not affect cosmic superstrings. However, it has been
argued \cite{pulsar} that with the full Parkes Pulsar Timing Array
(PPTA) project the upper bound will become $G\mu\leq 5\times
10^{-12}$, which is directly relevant for cosmic superstrings. In
conclusion, the full PPTA will either detect gravity waves from
strings and string-like objects, or they will rule out a number of
models.

\section{Conclusions}
\label{sec:concl}
Cosmic strings are generically formed at the end of hybrid inflation
in a large number of models within supersymmetry and supergarvity
theories.  String-like objects, which could play the r\^ole of cosmic
strings, are also generically produced at the end of brane inflation,
in many brane inflation models in the context of theories with large
extra dimensions. These one-dimensional objects would contribute in the
generation of fluctuations leading to the observed structure formation
and the measured CMB temperature anisotropies. They would also source
a stochastic gravity wave background.

Current measurements of the CMB spectrum, as well as of the
gravitational wave background, impose severe constraints on the free
parameters of the models. More precisely, the dimensionless parameter
$G\mu$ must be small enough to avoid contradiction with the currently
available data. 

The r\^ole of strings can be suppressed by adding new terms in the
superpotential \cite{mcdonald}, or by considering the curvaton
mechanism \cite{prl2005,endoetal}.  One can escape the {\sl string
problem} by complicating the models so that the produced strings
(D-term strings formed at the end of D-term inflation, or D-strings
formed at the end of brane collisions) become unstable (semilocal
strings), along the lines of Refs.~\cite{toine1,semilocal}. 
To be more specific, it has been proposed \cite{semilocal} that
introducing additional matter multiplets one obtains a
nontrivial global symmetry such as SU(2), leading to a simply
connected vacuum manifold and the production of semilocal
strings. Later on, it has been 
suggested \cite{toine1} that
if the waterfall Higgs fields are non-trivially charged under some
other gauge symmetries $H$, such that the vacuum manifold,
$[H\times U(1)]/ U(1)$, is simply connected, then the strings are
semilocal objects.

If the daily improved data require even more severe fine tuning of the
models, then I believe that one should develop and subsequently
study models where the strings
and string-like objects, formed at the end or after inflation,
are indeed unstable.

%
%

%
%



\end{document}